\documentclass[12pt]{article}
\pdfoutput=1
\usepackage[latin1]{inputenc}
\usepackage[title]{appendix}
\usepackage{amsfonts}
\usepackage{amssymb}
\usepackage{mathtools}
\usepackage{graphicx}
\usepackage{wrapfig}
\usepackage{caption}
\usepackage{color}
\usepackage{url}
\usepackage[colorlinks,linkcolor=black]{hyperref}

\makeatletter

\newcommand*\wthelper[2]{%
        \hbox{\dimen@\accentfontxheight#1%
                \accentfontxheight#11.15\dimen@
                $\m@th#1\widetilde{#2}$%
                \accentfontxheight#1\dimen@
        }%
}

\newcommand*\accentfontxheight[1]{%
        \fontdimen5\ifx#1\displaystyle
                \textfont
        \else\ifx#1\textstyle
                \textfont
        \else\ifx#1\scriptstyle
                \scriptfont
        \else
                \scriptscriptfont
        \fi\fi\fi3
}
\makeatother

\newcommand*{\sgn}{\ensuremath{\mathrm{sgn}}}

\begin{document}
\unitlength = 1mm

\title{\bf Duality Web on a 3D Euclidean Lattice \\ and Manifestation of Hidden Symmetries}

\author{Jun Ho Son$^1$, Jing-Yuan Chen$^1$ and S. Raghu$^{1,2}$\\
[7mm] \\
{\normalsize \it $^{1}$Stanford Institute for Theoretical Physics, Stanford 
University, Stanford, CA 94305, USA} \\
{\normalsize \it $^2$SLAC National Accelerator Laboratory, 2575 Sand Hill Road, 
Menlo Park, CA 94025, USA }\\
}

\maketitle

\begin{abstract}
We generalize our previous lattice construction of the abelian bosonization duality in $2+1$ dimensions to the entire web of dualities as well as the $N_f=2$ self-duality, via the lattice implementation of a set of modular transformations in the theory space. The microscopic construction provides explicit operator mappings, and allows the manifestation of some hidden symmetries. It also exposes certain caveats and implicit assumptions beneath the usual application of the modular transformations to generate the web of dualities. Finally, we make brief comments on the non-relativistic limit of the dualities.
\end{abstract}

\newpage

\tableofcontents

\section{Introduction}

 The notion of duality, namely that  two seemingly different quantum field theories are in fact descriptions of the same physical system, is a powerful concept with far-reaching consequences.  
 Some dualities relate strongly-coupled theories to weakly-coupled ones, allowing one to study strongly coupled systems perturbatively in weakly-coupled dual descriptions. Dualities between two strongly coupled theories do not endow one with such advantage, but they still shed light on aspects such as hidden global symmetries and relevant deformations that are not obvious from looking at only one side. In three spacetime dimensions, dualities have not only deepened our understanding on the non-perturbative aspects of strongly-coupled theories but are also  intimately related to the physics of surface states of strongly interacting topological insulators \cite{Metlitski2016,Wang2015}, superconductor-insulator transitions \cite{Mulligan2017}, the half-filled landau level \cite{Son2015}, and superuniversality in quantum Hall transitions \cite{Hui2018,Hui2018_2}, to name just a few.
 
Historically,  the fact that in three spacetime dimensions, a vector and second rank antisymmetric tensor (more generally a 1-form and 2-form respectively) have the same number of components, has served as an important motif for dualities.   This allows an operator mapping between matter currents and field strengths of a dual $U(1)$ gauge field. This mapping is a key ingredient in the famous boson-vortex duality \cite{Peskin1977,Dasgupta1981} and in Son's fermion-fermion duality \cite{Son2015}. The more recently conjectured dualities involve Chern-Simons-matter theories: various evidence 
indicates that there are a variety of dualities between two quantum field theories, one with fermionic matter and another with bosonic matter, coupled to Chern-Simons gauge fields, hence bearing the name ``bosonization'' dualities \cite{Chen1993,Burgess1994,Fradkin1994,Intriligator1996, Kapustin1999,Barkeshli2014,Kachru2015,Kachru2016,Aharony2016,Seiberg2016,Karch2016,Karch2017,Benini2018, Senthil:2018cru,Goldman2018}. Bosonization dualities makes earlier ideas on statistical transmutation in the context of the quantum Hall effect \cite{Arovas1984,Polyakov1988, zhang1989,Jain1989,lopez1991,Frohlich1991,wen1995,simon1998} more precise and extends them to gapless (i.e. critical) systems.  
 
The dualities of interest in this paper are dualities that follow from the simplest bosonization duality. The bosonization duality of interest here states that the following two Lagrangians (written in Euclidean signature) flow to the same IR fixed point:
\begin{eqnarray}
\label{eq:bfdual}
 -\mathcal{L}_{{\rm boson}}^{r}[A] &=&  -|(\partial_\mu-ib_\mu) \phi|^2 - r|\phi|^2 - \lambda |\phi|^4 + \frac{i}{4\pi} (A-b) d (A-b) \nonumber \\[.2cm]
& \updownarrow& \nonumber \\ -\mathcal{L}_{\rm Dirac}^{+, m}[A] &=& \bar\psi \, \sigma^\mu (\partial_\mu - i A_\mu) \, \psi + m\bar\psi\psi + \frac{i}{8\pi} A d A.
\end{eqnarray}
Here, $\phi$ and $\psi$ are, respectively, a complex boson field and two-component Dirac spinor, and the double arrow indicates that the two theories are dual to one another. $A$ and $b$ refer to the background electromagnetic field and a fluctuating $U(1)$ gauge field respectively. The duality is supposed to hold in the gapped phases with $\sgn(r)=\sgn(m)$, and most remarkably, also at the gapless critical point $r=0, m=0$. Lately, it was proposed that the $\mathcal{S}$ and $\mathcal{T}$ ``modular'' transformations, which explicitly modifies Lagrangians on the both sides, generate an infinite web of dualities starting from the seed duality Eq.~\eqref{eq:bfdual} \cite{Seiberg2016,Karch2016}. The $\mathcal{S}$ and $\mathcal{T}$ transformations were employed in describing the global phase diagram of the quantum Hall effect \cite{Kivelson1992}, and are defined as the following \cite{Witten:2003ya}:
\begin{equation}
\label{eq:modtrans}
\begin{split}
&\mathcal{S} : -\mathcal{L}[A] \rightarrow -\mathcal{L}[a] + \frac{i}{2\pi} a d B \\
&\mathcal{T} : -\mathcal{L}[A] \rightarrow -\mathcal{L}[A] + \frac{i}{4\pi} A d A.
\end{split}
\end{equation}
$\mathcal{S}$ can be understood as promoting the classical background gauge field $A$ to the quantum fluctuating gauge field $a$.  
Meanwhile, $\mathcal{T}$ can be simply interpreted as ``adding a Landau level''. $\mathcal{S}$ and $\mathcal{T}$ transformations generate $SL(2,\mathbb{Z})$ group, hence the name ``modular''. Remarkably, upon considering time-reversal transformation in addition to $SL(2,\mathbb{Z})$ transformations, one can ``derive'' the boson-vortex duality and the fermion-fermion duality from this web. The duality web hence bridges the two seemingly disparate classes of three-dimensional dualities.

While the aforementioned dualities pass many non-trivial checks, directly proving these dualities is extremely challenging. Take the seed bosonization duality, Eq.~\eqref{eq:bfdual},  for instance. The fermion side is free, but the boson side is strongly coupled, both due to the $|\phi|^4$ interaction and the 
strong gauge interaction. The duality makes statement about strongly-coupled theory without any special symmetries or structures that allow one to employ conventional analytical tools. Recently, there have been microscopic constructions that demonstrate the bosonization duality, one as an exact UV mapping between two partition functions in the Euclidean lattice \cite{Chen2018} and another in a wire array \cite{Mross2016,Mross2017}. These methods provide explicit physical insight into the nature of the operators involved in the duality mapping. In this paper, we will focus on the Euclidean lattice construction. The goal of this paper is to extend the Euclidean lattice construction of the bosonization duality Eq.~\eqref{eq:bfdual} to other dualities in the duality web. Some of the dualities in the web contains hidden global symmetries, and our lattice construction will make its origin manifest.

There is another issue we want to make manifest with our approach. It has sometimes been advocated that, if the seed duality Eq.~\eqref{eq:bfdual} could be rigorously proven, then the entire duality web generated by the $\mathcal{S}$ and $\mathcal{T}$ transformations would automatically follow. This cannot be taken for granted. While the $\mathcal{T}$ transformation involving background fields only is innocent, additional caveat underlies the $\mathcal{S}$ transformation. Both in the seed duality Eq.~\eqref{eq:bfdual} and the definition of the $\mathcal{S}$ transformation, to properly define the strongly coupled dynamical gauge field, we must regularize the theory in the UV. In this continuum theory this is usually done by suppressing its short-distance fluctuations by a Maxwell term. However, once this is taken into account, $\mathcal{S}$ and $\mathcal{T}$ do not exactly generate an $SL(2,\mathbb{Z})$ group structure; that the $SL(2,\mathbb{Z})$ is a good approximation in the strongly coupled IR should be viewed as part of the conjectural proposal. While this point is usually disguised and overlooked in the continuum presentation, in our microscopic construction its exposure is inevitable.

The paper is organized as follows. In Sec.~\ref{sec:dweb}, we explicitly show how to generate the duality web on the lattice, in particular, the implementation of $\mathcal{S}$ and $\mathcal{T}$ transformation on the Euclidean lattice. We will observe how our lattice construction captures the hidden quantum time-reversal symmetry in the duality web. We also discuss subtleties arising from adding Maxwell terms in the lattice partition function; these subtleties, as we will allude, are present in the continuum language as well though for slightly different reasons. In Sec.~\ref{sec:nf2}, we focus on the lattice UV completion of the $N_{f}=2$ duality. We discuss non-relativistic limit in Sec.~\ref{sec:nrlimit}. Concluding remarks appear in Sec.~\ref{sec:conc}

\section{Generating the Duality Web on Lattice}
\label{sec:dweb}

\subsection{Review of the Duality Web Structure}
\label{sec:dweb_cont}

In this section, we briefly show how the combination of modular $\mathcal{S}$, $\mathcal{T}$ and the time-reversal transformations realize other familiar dualities -- particularly, 1) the ``fermionization'' duality, 2) Son's fermion-vortex duality, and 3) the boson-vortex duality -- as part of the duality web. This section is intended to be a review, and our discussion will use continuum field theory; the lattice construction of the same structure, which is the main topic of this paper, will be discussed in Sec.~\ref{sec:trsmo}. 
 
Let's introduce some notation. We will work in the Euclidean signature. We abbreviate the $|\phi|^{4}$ theory coupled to a $U(1)$ gauge field $B$ as $-\mathcal{L}_{\rm WF}[B]$ ($B$ may be both classical or dynamical, depending on the situation. We reserve lower case roman letters for dynamical gauge fields, and capitals for classical, background gauge fields):
\begin{equation}
-\mathcal{L}_{\rm WF}^r[B] =  -|(\partial_\mu-iB_\mu) \phi|^2 - r|\phi|^2 - \lambda |\phi|^4.
\end{equation}
On the other hand, $\mathcal{L}_{\rm Dirac}^{+, m}[A]$ and $\mathcal{L}_{\rm Dirac}^{-, m}[A]$ denote Lagrangians of Dirac fermions coupled to a spin-$c$ connection $A$,
\footnote{In three spacetime dimensions, a spin-$c$ connection $A$ is much like a $U(1)$ connection, but with the extra property that, if we simultaneously change a fermion boundary condition and change the holonomy of $A$ across that boundary by $\pi$, the theory is left invariant. (In higher dimensions one must use a more involved definition; in three dimensions it can be formulated this way because all oriented three-manifolds admit spin structure.) The distinction of spin-$c$ versus ordinary $U(1)$ connection is useful for keeping track of the consistency of the theory \cite{Seiberg2016}; in this paper we will not use further details beyond the consistency check. In this paper, we always reserve $A, a$ for background / dynamical spin-$c$ connection while $B, b$ for background / dynamical $U(1)$ connection.}
with the sign of the superscript denoting the sign of the $\frac{i}{8\pi} a d a$ term that represents the parity anomaly (which is needed for invariance under large gauge transformations):
\begin{equation}
\begin{split}
-\mathcal{L}_{\rm Dirac}^{+, m}[A] &= \bar\psi \, \sigma^\mu (\partial_\mu - i A_\mu) \, \psi + m\bar\psi\psi + \frac{i}{8\pi} A d A \\
-\mathcal{L}_{\rm Dirac}^{-, m}[A] &= \bar\psi \, \sigma^\mu (\partial_\mu - i A_\mu) \, \psi + m\bar\psi\psi - \frac{i}{8\pi} A d A.
\end{split}
\end{equation}
The duality Eq.~\eqref{eq:bfdual} can therefore be written as:
\begin{equation}
\label{eq:bfdual1}
\begin{split}
&-\mathcal{L}_{{\rm boson}}[A] = -\mathcal{L}_{\rm WF}^r[b]  + \frac{i}{4\pi} (A-b) d (A-b) \\
& \updownarrow  \\ & -\mathcal{L}_{\rm Dirac}^{+, m}[A].
\end{split}
\end{equation}
In expressing the dualities, $\sgn(r)=\sgn(m)$ is always understood; the critical point, which is  of crucial interest, occurs when $r=0$, $m=0$.

Given a duality, the application of $\mathcal{S}$ or $\mathcal{T}$ to both sides generates new dualities. Let's start with an illuminating example. Applying $\mathcal{T}^{-1}\mathcal{S}\mathcal{T}^{-1}$ to both sides of Eq.~\eqref{eq:bfdual1} generates the duality between the following two theories:
\begin{eqnarray}
\label{eq:fbdual}
 -\mathcal{L'}_{\rm boson}[B] &=& -\mathcal{L}_{\rm WF}^r[b] + \frac{i}{4\pi} (a-b) d (a-b) - \frac{i}{4\pi} (a-B) d (a-B) \nonumber \\[.2cm]
& \updownarrow& \nonumber \\ -\mathcal{L}_{{\rm fermion}}[B] &=& -\mathcal{L}_{\rm Dirac}^{+, m}[a] - \frac{i}{4\pi} (a-B) d (a-B).
\end{eqnarray}
As for the boson side, expanding the last two Chern-Simons terms gives the following:
\begin{equation}
-\mathcal{L'}_{\rm boson}[B] = -\mathcal{L}_{\rm WF}^r[b] + \frac{i}{4\pi} b db - \frac{i}{4\pi} B dB + \frac{i}{2\pi} a d(B-b).
\end{equation}
Note that after the expansion, the gauge field $a$ appears nowhere but as a BF coupling with $(B-b)$. Thus, $a$ can be integrated out, acting as a Lagrange multiplier that enforces $B=b$. We find $\mathcal{L'}_{E, \, {\rm boson}}[B]$ is equivalent to $-\mathcal{L}_{\rm WF}^r[B]$, and Eq.~\eqref{eq:fbdual} says
\begin{eqnarray}
\label{eq:fbdual1}
& -\mathcal{L}_{\rm WF}^r[B] & \nonumber \\[.2cm]
& \updownarrow& \nonumber \\ & -\mathcal{L}_{{\rm fermion}}[B] \ = & -\mathcal{L}_{\rm Dirac}^{+, m}[a] - \frac{i}{4\pi} (a-B) d (a-B).
\end{eqnarray}
Compared to \eqref{eq:bfdual1}, we have essentially ``moved the Chern-Simons term" from the boson side to the fermion side, and this may be viewed as a ``fermionization'' duality. Sometimes one would write
\begin{equation}
-\mathcal{L}_{{\rm fermion}}[B] = -\mathcal{L}_{\rm Dirac}^{-, m}[a] + \frac{i}{2\pi} a d B - \frac{i}{4\pi} B d B.
\end{equation}
using the fact that $-\mathcal{L}_{\rm Dirac}^{\pm, m}[a]$ differ by a level $1$ Chern-Simons $\frac{i}{4\pi}ada$. This theory is commonly referred as ``Dirac fermion coupled to U$(1)_{-1/2}$ Chern-Simons gauge field', due to the half-level Chern-Simons term for the gauge field $a$. 

Note that in \eqref{eq:fbdual1}, $-\mathcal{L}_{\rm WF}^r[B]$ is manifestly time-reversal symmetric, while $-\mathcal{L}_{{\rm fermion}}$ is not. Thus, for the duality statement Eq.~\eqref{eq:fbdual1} to be valid, $-\mathcal{L}_{{\rm fermion}}$ must have an time-reversed version dual to itself in the IR. The time-reversed version of is denoted $-\mathcal{L}_{{\rm fermion}}^{T}$ or ``Dirac fermion coupled to U$(1)_{+1/2}$ Chern-Simons gauge field'':
\begin{equation}
\label{eq:Tu1f}
\begin{split}
-\mathcal{L}_{{\rm fermion}}^{T} = -\mathcal{L}_{\rm Dirac}^{-, -m}[a] + \frac{i}{4 \pi} (a-B) d (a-B) = -\mathcal{L}_{\rm Dirac}^{+, -m}[a] - \frac{i}{2\pi} a d B + \frac{i}{4\pi} B d B
\end{split}
\end{equation}
It is important to note that the mass has flipped sign. Thus, \eqref{eq:fbdual1} and its time-reversed version gives the IR dualities between three theories:
\begin{equation}
\label{eq:fbf}
-\mathcal{L}_{{\rm fermion}}[B] \ \longleftrightarrow \ -\mathcal{L}_{\rm WF}^r[B] \ \longleftrightarrow \ -\mathcal{L}_{{\rm fermion}}^{T}[B]
\end{equation}
where in this picture time-reversal refers to a left-right flip.

Son's fermion-fermion duality can be recovered from the aforementioned duality between $-\mathcal{L}_{{\rm fermion}}^{T}$ and $-\mathcal{L}_{{\rm fermion}}$. Think of applying $\mathcal{T}\mathcal{S}^{-1}\mathcal{T}$ transformation to Eq.~\eqref{eq:fbf}: Note that this is precisely the inverse transformation of the operation we applied on Eq.~\eqref{eq:bfdual1} to generate the duality Eq.~\eqref{eq:fbdual1}. Thus, applying $\mathcal{T}\mathcal{S}^{-1}\mathcal{T}$ on $-\mathcal{L}_{{\rm fermion}}$ and $-\mathcal{L}_{\rm WF}^r[B]$, we recover the original theories in Eq.~\eqref{eq:bfdual1}. On the other hand, action of $\mathcal{T}\mathcal{S}^{-1}\mathcal{T}$ on $-\mathcal{L}_{{\rm fermion}}^{T}$ does not give a free theory. Instead, the new action $-\mathcal{L}_{\rm QED}$ is strongly-interacting:
\begin{equation}
\begin{split}
-\mathcal{L}_{\rm QED}[A] &= -\mathcal{L}_{\rm Dirac}^{-, -m}[a] + \frac{i}{4 \pi} (a-b) d (a-b) + \frac{i}{4 \pi} (A-b) d (A-b) \\
&= -\mathcal{L}_{\rm Dirac}^{+, -m}[a] + \frac{i}{2 \pi} b d b + \frac{i}{4 \pi} A d A - \frac{i}{2 \pi} b d (A+a)
\end{split}
\end{equation}
The above theory is known as the properly quantized version of QED$_{3}$. The duality between the free Dirac fermion $-\mathcal{L}_{\rm Dirac}^{+, m}[A]$ and the QED$_{3}$ that we just derived, is Son's fermion-vortex duality. The more familiar ``illegal" version of $\mathcal{L}_{\rm QED}$ in the condensed-matter literature can be recovered by integrating out $b$ -- however, there is level-$2$ Chern-Simons term for $b$, so integrating out $b$ cannot be justified on general grounds, and yields a loose, local expression with improperly quantized topological terms.

On the other hand, applying $\mathcal{T}^{-1} \mathcal{S} \mathcal{T}^{-1}$ transformation to Eq.~\eqref{eq:fbf}, we obtain the time-reversed version of the theories in Eq.~\eqref{eq:bfdual1} as well as that of $-\mathcal{L}_{\rm QED}$, and hence a time-reversed version of Son's duality.

Finally, we will see how the more familiar boson-vortex duality can be recovered from the duality web. Note that \eqref{eq:Tu1f}, which is dual to $-\mathcal{L}_{\rm WF}^r[B]$, has two equivalent expressions. However, the latter expression may be viewed as applying $\mathcal{T}\mathcal{S}^{-1}$ to the fermion side of Eq.~\eqref{eq:bfdual1} (but with $-m$ in place of $m$), and therefore is dual to applying $\mathcal{T}\mathcal{S}^{-1}$ to the boson side of Eq.~\eqref{eq:bfdual1} (but with $-r$ in place of $r$): 
\begin{equation}
\begin{split}
-\mathcal{L}_{\rm AbHiggs}[B] &= -\mathcal{L}_{\rm WF}^{-r}[b]  + \frac{i}{4\pi} (a-b) d (a-b) - \frac{i}{2\pi} a d B + \frac{i}{4\pi} B d B  \\
&= -\mathcal{L}_{\rm WF}^{-r}[b] - \frac{i}{2\pi} b d B + \frac{i}{4\pi} (a-b-B) d (a-b-B).
\end{split}
\end{equation}
The $\frac{i}{4\pi} (a-b-B) d (a-b-B)$ is a level-$1$ Chern-Simons term whose dynamics is completely decoupled from other parts of the Lagrangian, so this term can be integrated out. Upon doing this, we arrive at the Lagrangian for the Abelian Higgs model. Hence, we observed that the boson-vortex duality, $-\mathcal{L}_{\rm WF}^r[B] \leftrightarrow -\mathcal{L}_{\rm AbHiggs}[B]$ (note that the Wilson-Fisher theory and the abelian Higgs theory have opposite signs of $r$), is part of the duality web as well.

\begin{figure}
\includegraphics[width=\textwidth]{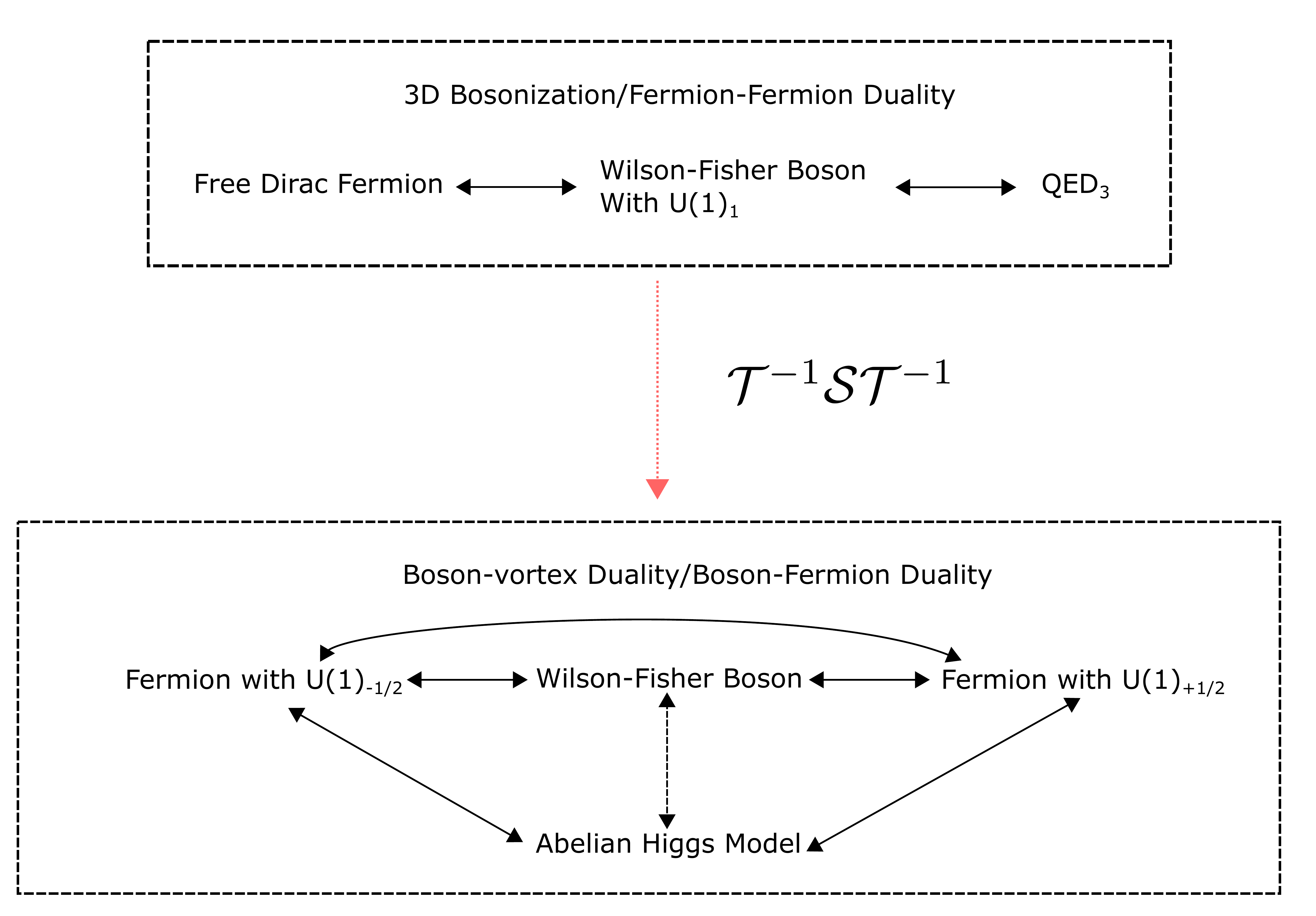}
\caption{Overview of dualities and their relations to each other. Theories in the dashed box are conjectured to be dual to each other; the modular transformation $\mathcal{T}^{-1}\mathcal{S}\mathcal{T}^{-1}$ drawn as a dotted red arrow allows one to generate theories on one dashed box from the theories in the different box. We present dualities marked as lines with both arrowheads as exact lattice dualities in this paper. Boson-vortex duality, marked as a dashed line with both arrowheads, cannot be expressed as exact lattice mapping in our paper.}
\label{fig:summary}
\end{figure}

We summarized the relation between dualities mentioned in this section in Fig.~\ref{fig:summary}. In \cite{Chen2018}, we presented exact lattice partition function mappings that can be viewed as UV demonstration of the 3D bosonization duality Eq.~\eqref{eq:bfdual1}, by implementing Chern-Simons terms with heavy fermions. Following the same ideas, in this paper we realize dualities in Fig.~\ref{fig:summary} as UV lattice mappings as well. We will show that such UV lattice mappings exhibit non-trivial properties anticipated from the continuum viewpoint as well. While we could not write down the boson-vortex duality as an exact lattice mapping, we will show within our formalism that this duality does exist as an IR duality.
 
Before moving onto the lattice construction, we make some comments on the inclusion of Maxwell terms to regularize the continuum field theories presented in this section. We first focus on the boson side of the seed bosonization duality, i.e., $\mathcal{L}_{{\rm boson}}[B]$ of Eq.~\eqref{eq:bfdual1}. This theory involves a fluctuating gauge field $b$ coupled to the boson, so it is a strongly coupled quantum field theory.  Since the gauge charge $e^2$ runs strong in the IR in 3 spacetime dimensions, a strongly coupled theory is a natural expectation at low energies. To define such a theory perturbatively in the UV, one may regularize the theory with a UV cutoff $\Lambda_{UV}$, and add a Maxwell term $-\frac{(db)^2}{4e^2}$ to the Lagrangian, with the choice $e^2 \ll \Lambda_{UV}$.  Nonetheless, for energy scales $\mu$ of interest, namely $\mu \ll e^2 \ll \Lambda_{UV}$, the desired behavior will emerge. 
 
Similar issues arise in the application of the $\mathcal{S}$ transformation as well. Since $\mathcal{S}$ turns the background gauge field into the fluctuating gauge field, for the theory to remain well-defined in the UV after the transformation, Maxwell terms for the gauge fields should be supplemented to suppress UV modes of the  now fluctuating gauge fields. For example, $\mathcal{L'}_{E, \, {\rm boson}}[A]$ of Eq.~\eqref{eq:fbdual} should be modified as the following:
\begin{equation}
-\mathcal{L'}_{\rm boson}[B] \rightarrow -\mathcal{L'}_{\rm boson}[B] - \frac{(da)^2}{4e_{a}^2} - \frac{(db)^2}{4e_{b}^2} 
\end{equation}
Once more, 
$\mu \ll e_{a}^2, e_{b}^2 \ll \Lambda_{\text{UV}}$. However, it turns out that for the now-regularized theory to mimic  similar IR behavior as one expects from the naive unregularized picture, one should be able to integrate out $a$ to impose constraint $B=b$. Note that the Lagrangian is still quadratic in $a$, and after integrating out $a$, one obtains
\begin{equation}
-\mathcal{L}_{\rm WF}[b] + \frac{i}{4\pi} b db - \frac{i}{4\pi} B dB - \frac{e_{a}^{2}(B-b)^2}{16\pi^{2}} - \frac{(db)^2}{4e_{b}^2} 
\end{equation}
Note that the delta function $\delta(B-b)$ obtained from integrating out $a$ is now ``softened" to the superconducting mass term $- \frac{e_{a}^{2}(B-b)^2}{16\pi^{2}}$ (in the London gauge); sending $e_{a} \rightarrow \infty$ enables us to recover the delta function. In order for the Lagrangian to truly represent the  Wilson-Fisher fixed point given by $\phi^{4}$ theory, $e_{a}^2$ should be a large parameter at the energy scale where the gauge field $b$ may acquire non-trivial dynamics, i.e., $e_{b}^2 \ll e_{a}^2$. Thus, the strict hierarchy 
$\mu \ll e_{b}^2 \ll e_{a}^2 \ll \Lambda_{\text{UV}}$ should be enforced to make UV of the quantum field theory well-defined, and at the same time, to retain the IR  physics of the boson side of Eq.~\eqref{eq:fbdual}.   

However, now there is a caveat. Suppose we want to interpret the boson side of Eq.~\eqref{eq:fbdual} using the fermion side instead. For that purpose, we would like to invoke the IR duality Eq.~\eqref{eq:bfdual1} first \emph{before} we integrate out $a$. That is, we would want $e_a^2 \ll e_b^2$ so that we can integrate out the $b$ gauge field first to obtain the Dirac Lagrangian, which is contrary to the condition above. Therefore, to ``generate Eq.~\eqref{eq:fbdual1} from Eq.~\eqref{eq:bfdual1}'', on the boson side we assumed $e_b^2 \ll e_a^2$, while on the fermion side we assumed $e_a^2 \ll e_b^2$. We have therefore implicitly made an assumption that changing the relative strengths of $e_b$ and $e_a$ over some finite range of energies does not alter the IR physics significantly -- including the physics at the critical point. More compactly, such implicit assumption underlies the seemly innocent rule ``$\mathcal{S} \mathcal{S}^{-1} = \mathbf{1}$'' once the $\mathcal{S}$ and $\mathcal{S}^{-1}$ are regularized by Maxwell terms; in other words, it underlies the recognition of $\mathcal{S}$ and $\mathcal{T}$ as ``generating'' the algebra of $SL(2, \mathbb{Z})$ to a good approximation in the IR. This implicit assumption has not been emphasized in the literature, but it will inevitably become manifest in our lattice construction.

\subsection{Review of the Euclidean Lattice Construction of the Bosonization Duality}

 For the readers who are not familiar with our Euclidean lattice version of the bosonization duality \cite{Chen2018} and also for the sake of setting up the notation that carries throughout the rest of the paper, here we briefly review the construction presented in our previous work. 
 
 Our construction involves a lattice gauge theory constructed on the 3D cubic lattice. The two main ingredients of our construction are the XY model lattice action
\begin{equation}
\label{eq:XY}
Z_{\rm{XY}}[A;T] =  \prod_{n} \left( \int^{\pi}_{-\pi} \frac{d \theta_{n}}{2\pi} \right) e^{-H_{\rm{XY}}[\theta,A ; T]}, \, -H_{\rm{XY}}[\theta,A ; T] = \frac{1}{T} \sum_{n\mu} \cos{(\theta_{n+\hat{\mu}}-\theta_{n} - A_{n\mu})}
\end{equation} 
and the Wilson fermion action
\begin{equation}
\label{eq:WF}
\begin{split}
&Z_{W}[A;M,U] = \int D\bar{\psi}D\psi = \prod_{n} \left( \int d\bar{\psi}_{n} d\psi_{n}\right) e^{-H_{\rm{W}}[\bar{\psi},\psi,A ; M, U ]} \\
-&H_{\rm{W}}[\bar{\psi},\psi,A ; M, U ] = \sum_{n\mu} \bar{\psi}_{n+\hat{\mu}}e^{i A_{n\mu}}\frac{\sigma^{\mu}-1}{2} \psi_{n} + \bar{\psi}_{n}e^{-i A_{n\mu}}\frac{-\sigma^{\mu}-1}{2} \psi_{n+\hat{\mu}} \\
& \hspace{4cm} + M \sum_{n} \bar{\psi}_{n}\psi_{n} + U \sum_{n\mu} \bar{\psi}_{n+\hat{\mu}}\frac{\sigma^{\mu}-1}{2} \psi_{n} \bar{\psi}_{n}\frac{-\sigma^{\mu}-1}{2} \psi_{n+\hat{\mu}} .
\end{split}
\end{equation}
In the expression above, $n$ is an index that labels each lattice site.  Unit vectors that span the three-dimensional cubic lattice are labeled by $\hat{\mu} = \hat{x}, \hat{y}, \hat{z}$; hence, $n+\hat{\mu}$ refers to a site neighboring $n$. $n\mu$ is an index associated with a link pointing from site $n$ to site $n+\hat{\mu}$. $\sigma^{\mu}$ refers to $2 \times 2$ pauli matrices.
Matter fields such as two-component Grassmann fields $\psi$ and $\chi$,  and the boson phase field $\theta$ (a $U(1)$ compact variable) live on lattice sites. Meanwhile, the $U(1)$ gauge fields $A$ and $b$, which are also compact, reside on the links of the lattice.   

In the action of Eq.~\eqref{eq:XY}, tuning $T$ induces a  phase transition from a trivially gapped phase to a superconducting phase in which $A$ is higgsed. The critical point between two phases is a lattice realization of the O$(2)$ Wilson-Fisher fixed point. Meanwhile, the action in Eq.~\eqref{eq:WF} corresponds to two-component free Dirac fermions supplemented with four-fermi interaction $U$ that takes a role of a ``counterterm" in the exact mapping. When $U=0$, one can explicitly compute the ``bubble diagram" (the leading contribution to the gauge field self-energy) to see that the level-$k$ Chern-Simons term for the gauge field $A$ is generated; $k$ takes different integer values according to the value of $M$, as the following:
\begin{equation}
k = \begin{cases}
0 & |M|>3\\
1 & 1<|M|<3 \\
-2 & |M|<1
\end{cases}.
\end{equation}
The regime of our interest is $1<M<3$ --- since this regime corresponds to a $k=1$ gapped phase, one can understand IR lattice Chern-Simons action as generated from integrating out gapped fermion with proper $M$ value. Thus, the gapped fermion action can be used to implement a Chern-Simons term for the gauge field on the boson side of the duality Eq.~\eqref{eq:bfdual}.

 Another interesting point is $M=3$, the critical point between $k=1$ and $k=0$ phase. At $M=3$, one can check 
by expanding near $k=(0,0,0)$ that a single gapless Dirac cone emerges in IR; one can employ the lattice action Eq.~\eqref{eq:WF} near $M=3$ to implement the free-fermion side of the duality Eq.~\eqref{eq:bfdual}, with $m\simeq M-3$. Finally, we stress that four-fermi interaction is irrelevant on the ``free-fermion line'' $U=0$. Turning on small $U$ term may renormalize the critical value of $M$ slightly from $3$ by order $U$, yet qualitative features of free-fermion phase diagram remain valid in small but finite $U$. (Although $M=1$ is also a critical point, it is one across which the Chern-Simons level $k$ jumps by $3$, and hence not generic. It is an artifact of the use of cubic lattice, so it does not describe universal physics and we do not focus on it.)

Given these two main ingredients, upon choosing proper parameters, one can expect two partition functions defined on the lattice, $Z_{B}$ and $Z_{F}$, to be the lattice realization of boson and fermion side of the bosonization duality Eq.~\eqref{eq:bfdual1} respectively:
\begin{equation}
\label{eq:bosonizationpt}
\begin{split}
&Z_{B}[A;T,M,U] = \int Db \, Z_{\rm{XY}}[b;T]Z_{W}[A-b;M,U] \\
&Z_{F}[A;M',U'] = Z_{W}[A;M',U'],
\end{split}
\end{equation}
where the integration measure is simply defined by
\begin{equation}
\int D b = \prod_{n\mu}\int^{\pi}_{-\pi} \frac{db_{n\mu}}{2\pi}.
\end{equation}
In particular, on the boson side, $1<M<3$ so that $Z_W$ implements level $1$ Chern-Simons, and the coupled theory has some critical $T_c$ that $r \sim T-T_c$. On the fermion side, $m \sim M'-3$ of the IR Dirac fermion. We also want both $U$ and $U'$ to be small so that we may safely use the fact that they are irrelevant.

Remarkably,  one can explicitly integrate out $\theta$ and $b$ field in $Z_{B}$ to show that
\begin{equation}
\label{eq:bosonizationduality}
Z_{B}[A;T,M,U] \propto Z_{F}[A;M',U'],
\end{equation}
where $M'$ and $U'$ are related to $T$, $M$ and $U$ as the following:
\begin{equation}
\label{eq:parameter_change}
M' = \frac{I_{0}(\frac{1}{T})}{I_{1}(\frac{1}{T})} M, \quad U' = \left( \frac{I_{0}(\frac{1}{T})}{I_{1}(\frac{1}{T})} \right)^2 (1+U) -1,
\end{equation}
Here $I_0$ and $I_1$ denote the modified Bessel functions -- their details do not matter, it only matters that their ratio monotonically increases from $1$ to infinite as $T$ increase from $0$ to infinite. Therefore, started with $1<M<3$, there must be some $T_c$ around which $M' \simeq 3$. Of course, the critical mass might not occur exactly at $M'=3$ because we must also take $U'$ into account. A perturbative calculation in $U$ and $U'$ (compared to the lattice scale) shows that there is a parameter regime in which $Z_{W}[A;M',U']$ can be interpreted as a single gapless Dirac fermion, and at the same time $Z_{W}[A-b;M,U]$ as a heavy Dirac fermion generating a level-1 Chern-Simons term. The critical temperature $T_c$ thus occurs at an order of $U'$. In addition to a partition function mapping, one can also show that there is the following relation between correlation functions, for any configuration of the background gauge field $A$:
\begin{equation}
\left\langle e^{i\theta_{n_1}}\chi_{n_1} \cdots e^{i\theta_{n_k}}\chi_{n_k} \, e^{-i\theta_{n'_1}}\bar\chi_{n'_1} \cdots e^{-i\theta_{n'_k}}\bar\chi_{n'_k} \right\rangle \propto \left\langle \psi_{n_1} \cdots \psi_{n_k} \bar\psi_{n'_1} \cdots \bar\psi_{n'_k} \right\rangle
\end{equation}
where $\chi$ and $\psi$ respectively denotes fermion field in $Z_{B}[A;T,M,U]$ and in $Z_{F}[A;M',U']$. Identifying $e^{\pm i\theta_{n}}$ as boson creation/annihilation operator and $\chi_{n}$ as monopole operators, this operator mapping fits nicely into continuum picture in which a composite object of a boson and a monopole forms a Dirac fermion. 

We make a final remark in relation to the Maxwell terms for the fluctuating gauge field $b$. In the continuum theory, we mentioned that a Maxwell term must be introduced to regularize the strongly coupled theory. In the lattice presentation, since the theory is already regularized, why do we still include a Maxwell term? First, it can be included, so we should include it for generality. But a better reason is that, in the partition function $Z_{B}[A;T,M,U]$ of Eq.~\eqref{eq:bosonizationpt}, the idea that integrating out the heavy Wilson fermion $Z_{W}[A-b;M,U]$ yields predominantly a  Chern-Simons term can only be justified if $A-b$ is a small fluctuation. If it fluctuates strongly the connection to the continuum formulation becomes unclear due to the presence of large irrelevant operators (which in turn imply that we are far from the fixed point described by the continuum theory). To suppress these UV modes, one may add the Maxwell term by hand. For example, one can imagine including the following, ``non-compact'' Maxwell \cite{Peskin1977}, into the partition function $Z_{B}[A;T,M,U]$:
\begin{equation}
\label{eq:Maxwell}
Z_{M}[b; g] = \left( \prod_{\rm plaq. \, \Box} \ \sum_{l_\Box \in \mathbb{Z}} \right) \ \left( \prod_{\rm cubes} \ \delta_{l \, \rm{closed}} \right) \ \exp{\left[ - \frac{1}{4 g^2} \sum_{ \Box} \left( (\vec \triangle \times \vec b)_\Box + 2\pi l_\Box \right)^2 \right]}
\end{equation}
where $\Box$ labels the lattice plaquettes, $\vec \triangle \times$ is the lattice curl, and $l_\Box$ is a \emph{closed} integer field on the plaquettes -- ``closed'' meaning that the sum of the $l$ field coming out of the faces of each lattice cube must vanish.
\footnote{Since the $l$ field is closed, locally it may also be written as the lattice curl of an integer gauge field on lattice links, $l=\vec\triangle\times\vec m$. The integer gauge field $m$ may then be combined with the $U(1)$ gauge field $b$ as $b+2\pi m$ which takes \emph{real} value, hence the name ``non-compact''. On the other hand, if we relax the closedness condition of the $l$ field, then $Z_M$ will be the Villainized version of the ``compact'' Maxwell term \cite{Peskin1977}, whose usual version has $-S=(1/2g^2) \sum_\Box \cos \left(\vec\triangle\times \vec b\right)_\Box$ in the exponent, without the $l$ field. In the Villainized ``compact'' Maxwell, the integer $l$ field is interpreted as the Dirac strings, whose end points (i.e. lattice cubes out of which $l$ is not closed) are fluctuating monopoles, so that the total flux $\vec \triangle \times \vec b + 2\pi l$ appears non-conserved. \label{compact_Maxwell}}  
Similar to the continuum case discussed previously, we set $g^2 \ll \text{(lattice scale)}$ to suppress UV modes and  
$\mu  \ll g^2$ to preserve non-trivial IR dynamics. Note the similarity to the requirements for $e^2$ in the bosonization duality in the previous section. Hence, in the lattice construction, the similar issue of including Maxwell terms we observed in the continuum formulation persists as well. However, the origin is very different: The lattice constructions are always UV-regularized, but \textit{the Maxwell terms are needed to make the connection to the continuum theory manifest}.
 
The current status \cite{Senthil:2018cru} of the microscopic construction is therefore that, without a Maxwell term (i.e. $g^2\rightarrow \infty$ on the lattice), we have an exact lattice duality mapping, but the connection to the continuum definition of theories (where $e^2$ is small in the UV) can only be argued based on the idea that $e^2$ in the continuum runs towards infinite in the IR anyways. Whilst, with a Maxwell term (i.e. $g^2$ small on lattice), the connection to the continuum theories is better justified, the exact duality map is absent. \footnote{Some relevant discussions on this point can be found in Appendix C of \cite{Chen:2018vmz}.} Later we will see that the implementation of the $\mathcal{S}$ transformation is under the same status, and the issue is related to the caveat that also exists in the continuum theory which we mentioned at the end of Sec.~\ref{sec:dweb_cont}.

%
%

\subsection{Lattice Presentation of Duality Web and Time-Reversal Symmetry}
\label{sec:trsmo}

Now, we will see how the above continuum picture of the duality web is realized on a lattice. The main idea is that the $\mathcal{T}$ operation in the continuum is to add a Chern-Simons term $\frac{i}{4\pi} AdA$; also the $\mathcal{T}^{-1}\mathcal{S}\mathcal{T}^{-1}$ operation can be thought of as promoting the background gauge field $B$ to a dynamical field $b$ and to add a Chern-Simons term $-\frac{i}{4\pi} (b-A)d(b-A)$. Together they generate all the $SL(2,\mathbb{Z})$ transformations. On the lattice, the addition of Chern-Simons term can be implemented by the heavy Wilson-fermion partition function $Z_{W}^{*}[b-A; M, U=0]$. ($*$ denotes complex conjugation)
\footnote{While $Z_W[A, M, 0]$ implements $-\mathcal{L}^{+, m}_{\rm Dirac}[A]$ with $m\simeq M-3$, its complex conjugate $Z^\ast_W[A, M, 0]$ implements $-\mathcal{L}^{-, -m}_{\rm Dirac}[-A]$. Moreover, $-\mathcal{L}^{-, -m}_{\rm Dirac}[-A]$ is equivalent to $-\mathcal{L}^{-, -m}_{\rm Dirac}[A]$ by a charge conjugation transformation.}.
For instance, from Eq.~\eqref{eq:bfdual1} to Eq.~\eqref{eq:fbdual}, we implement $\mathcal{T}^{-1}\mathcal{S}\mathcal{T}^{-1}$ operation on the lattice as the following:
\begin{equation}
\label{eq:fbduallat}
\begin{split}
&Z_{B}[A;T,M] \rightarrow Z_{B'}[B;T,M, M_{1}] = \int Da Db \, Z_{\rm{XY}}[b;T] Z_{W}[a-b;M,0] \ Z_{W}^{*}[a-B; M_{1}, 0]  \\
&Z_{F}[A;M',U'] \rightarrow Z_{F'}[B;M',U',M_{1}] = \int Da \, Z_{W}[a;M',U'] \ Z_{W}^{*}[a-B; M_{1}, 0]
\end{split}
\end{equation}
where $1<M_1<3$, and we have set the irrelevant coupling $U$ in Eq.~\eqref{eq:bosonizationpt} to zero for convenience. We stress that the equivalence between the two partition functions in the equation above is \textit{exact}, with $M'$ and $U'$ given in terms of $T$, $M$, $U=0$, as defined in Eq.~\eqref{eq:parameter_change}. 

In the continuum, one could integrate out the gauge field $a$ on the boson side of Eq.~\eqref{eq:fbdual} to see that the theory on the boson side is actually $\phi^{4}$ theory in disguise, if one implicitly assumes the Maxwell term can be neglected in an $\mathcal{S}$ transformation. However, such procedure of integrating out the gauge field \textit{cannot be done exactly on the lattice partition function} $Z_{B'}[B;T,M,M_{1}]$. An intuitive argument that this \emph{must} not be possible on the lattice is that, Maxwell and other irrelevant terms are necessarily generated along with the Chern-Simons term from the heavy Wilson fermion.

Let's formulate this in detail. Ignoring Maxwell terms, $(\mathcal{T}\mathcal{S}^{-1}\mathcal{T}) \, (\mathcal{T}^{-1}\mathcal{S}\mathcal{T}^{-1})$ is the identity, i.e. imposes the ``Lagrange multiplier constraint'' $\delta(b-B)$; with the Maxwell terms included, this delta function is softened to the Higgsing of $b-B$. We expect this on the lattice. Note that $Z_{B'}[B;T,M,M_{1}]$ in \eqref{eq:fbduallat} can be written as $\int Db \, Z_{\rm{XY}}[b;T] \: Z_{DW}[B,b;M,M_{1}]$ where $Z_{DW}$ is the partition function of a ``doubled" set of Wilson fermions and is given by
\begin{equation}
\label{eq:doublewf}
Z_{DW}[B,b;M,M_{1}] = \int Da \, Z_{W}[a-b;M,0] \ Z_{W}^{*}[a-B; M_{1}, 0].
\end{equation}
Thus, the above is a lattice implementation of $(\mathcal{T}\mathcal{S}^{-1}\mathcal{T}) \, (\mathcal{T}^{-1}\mathcal{S}\mathcal{T}^{-1})$. We expect this partition function, in appropriate range of parameters, to exhibit the Higgsing of $b-B$. Remarkably, $Z_{DW}$ can be simulated using Monte Carlo methods without a sign problem if we set $M = M_{1}$ (there is no obstruction to doing so) and if we turn off the gauge fields $b$ and $B$ -- it is sign problem free simply because the integrand of Eq.~\eqref{eq:doublewf} becomes manifestly positive. If we show $Z_{DW}$ at these parameters exhibits a 
superconducting phase, then it would Higgs $b-B$ at least for small values of $b-B$ (this is in parallel to our $e_b^2 \ll e_a^2$ discussion in the continuum). Indeed, in the numerical simulation -- whose details we present in the next subsection -- we show a fermion bilinear, with charge $1$ with respect to $B$ and charge $-1$ with respect to $b$, condenses in an extended range of $M$, supporting the idea that $\int Da \, Z_{W}[a-b;M,0]Z_{W}^{*}[a-B; M, 0]$ \textit{implements $\delta(B-b)$ in the IR of $Z'_{B}[B;T,M, M]$} for some choices of $M$ (the support is reliable at small fluctuations of $b$, as we will discuss later). This information about the partition function Eq.~\eqref{eq:doublewf} provides evidence that the partition function $Z_{B'}[B;T,M,M]$ and $Z_{\rm{XY}}[B;T']$ flows to the same IR fixed point for some proper choice of $M$. This gives us confidence that the exact correspondence between two partition functions in Eq.~\eqref{eq:fbduallat}, implementing Eq.~\eqref{eq:fbdual}, can indeed be further recognized as the implementation of Eq.~\eqref{eq:fbdual1}. The connections of the partition functions in Eq.~\eqref{eq:fbduallat} to continuum QFTs are more reliable upon including Maxwell terms; we reserve comments about this to the end of this subsection.

The above is the lattice realization of Eq.~\eqref{eq:fbdual1}, i.e. the left portion of Eq.~\eqref{eq:fbf}. Its time-reversed version, the right portion of Eq.~\eqref{eq:fbf}, also manifests as exact partition function mapping. To see this, in Eq.~\eqref{eq:fbduallat}, make the variable change
\begin{equation}
\label{eq:gfchange}
a = -a'+b+B
\end{equation}
on the partition function $Z_{B'}[B;T,M,M]$. Note that the above variable change can be implemented as an SL$(3,\mathbb{Z})$ transformation on the vector $(a,b,B)$, so it is a permissible variable change that preserves the compact nature of the gauge field. Then, one can write
\begin{equation}
Z_{B'}[B;T,M,M] = \int Da Db \, Z_{\rm{XY}}[b;T]Z_{W}[B-a';M,U]Z_{W}^{*}[b-a'; M_{1}, 0].
\end{equation}
Now we can integrate out $b$ above and find that it exactly maps to
\begin{equation}
\label{eq:timeRdualF}
Z_{F'}^{T}[B;M',U',M] = \int Da' \, Z_{W}^{*}[-a';M',U']Z_{W}[B-a'; M, 0].
\end{equation}
Note that in $Z_{F'}$, the role of the heavy fermion that simply generates the Chern-Simons term is played by $Z_{W}^{*}$; $Z_{W}$ plays the role of the light Dirac fermion, whose mass parameter drives the theory to the criticality. In $Z_{F'}^{T}[B;M',U',M]$, the roles of the heavy fermion and the light fermion \textit{reversed}, hence fitting the description ``time-reversed dual''. (Comparing our definitions of $Z_{F'}$ and $Z_{F'}^{T}$, one might note that in addition to switching the fermions, the sign of the gauge fields in $Z_W$ also flipped. But this does not matter because $Z_W[A]=Z_W[-A]$ by a charge-conjugation transformation.) This switch of roles simply follows from the two presentations of $Z_{B'}$ upon changing the variables Eq.~\eqref{eq:gfchange}.

Now that we have the lattice version of Eq.~\eqref{eq:fbf}, we may generate the lattice version of Son's duality as in the continuum. We implement the lattice $\mathcal{T}\mathcal{S}^{-1}\mathcal{T}$ on the action $Z_{F'}^{T}$ and $Z_{F'}$ by making $B \rightarrow b$ dynamical and add another heavy fermion $Z_{W}[A-b;M,0]$. This gives the two lattice actions:
\begin{equation}
\begin{split}
Z_{F''}[A;M',U',M] &= \int Da Db \, Z_{W}[a;M',U']Z_{W}^{*}[a-b; M_{1}, 0] Z_{W}[A-b;M,0] \\
&= \int Da \, Z_{W}[a;M',U'] Z_{DW}[-A,-a;M,M] \\
Z_{\rm QED}[A;M',U',M] &= \int Da Db \, Z_{W}^{*}[-a;M',U']Z_{W}[-a+b; M, 0] Z_{W}[A-b;M,0]
\end{split}
\end{equation}
Once again, as for the lattice action $Z_{F''}$, $Z_{DW}[-A,-a;M,M]$ is expected to Higgs $A-a$; thus, $Z_{F''}$ flows to the free-fermion $Z_W$ with $a=A$. However, $Z_{\rm QED}$ is expected to be interacting, and it is straightforward to recognize this theory as the lattice version of QED$_{3}$ that appears in fermion-fermion duality.
 
Finally, we comment on the inclusion of Maxwell terms. Consider the fermion side $Z_F$ in Eq.~\eqref{eq:fbduallat} first. For the heavy Wilson fermion factor $Z_{W}^{*}[a-B; M_{1}, 0]$ to be reliably interpreted as the Chern-Simons term, one should include Maxwell term Eq.~\eqref{eq:Maxwell} $Z_{M}[a;g_{a}]$ on both sides of the partition function. This Maxwell term plays a very similar role to the one played in the boson side of the lattice bosonization duality in the previous subsection. The coupling $g_{a}$ also follows a similar constraint of 
$\mu \ll g_{a}^{2} \ll \text{(lattice scale)}$. Note that adding this term \textit{does not} break the exact partition function mapping in Eq.~\eqref{eq:fbduallat}. Also, the theory Eq.~\eqref{eq:doublewf} can be still simulated without a sign problem upon including the Maxwell term \footnote{Including large Maxwell terms generically increase the correlation length, and finite-size effects, forcing one to investigate larger systems to make sense of the true thermodynamic behavior of the system. Hence, we reserve simulation of this theory upon including large Maxwell term to the future work.}.

On the other hand, let's consider the boson side $Z_B$ in Eq.~\eqref{eq:fbduallat}. We claimed its physics should realize that of $Z_{XY}$, because the $Z_{DW}$ factor Eq.~\eqref{eq:doublewf} Higgses out the field $b-B$. This Higgsing is reliable only for  small fluctuations of $\left( b-B \right)$. Therefore, to make it reliable, one needs to include the Maxwell term for the gauge field $b$ on the boson side $Z_b$. For $g_{b}$ to suppress dangerous UV large fluctuations, it should be much smaller than the scale encoded by lattice partition functions and $g_{a}$ but much larger than IR modes. Hence, the hierarchy 
$\mu \ll g_{b}^{2} \ll g_{a}^{2} \ll \text{(lattice scale)}$ should be satisfied to make connection to the boson side of the continuum field theory. Now a caveat similar to that at the end of Sec.~\ref{sec:dweb_cont} arises. Upon including Maxwell terms for $b$, the whole exact mapping from the boson side to the fermion side becomes unavailable, as mentioned at the end of the previous subsection. But for the fermion side to be sensibly interpreted as ``Dirac fermion plus Chern-Simons term'', we expect $g_a^2\ll g_b^2$, which is opposite to the range of parameters for the boson theory to become $Z_{XY}$. So we encounter the same situation as in the continuum, where different sides of the ``new'' duality requires different ranges of parameters, in order to have it ``derived'' from old dualities.

It is also worth noting that, upon the inclusion of the Maxwell terms for $a$ and $b$, the time reversal picture under Eq.~\eqref{eq:gfchange} becomes non-exact.

Once more, since \textit{not including Maxwell terms} make the non-trivial properties of the dualities manifest, it is natural to guess that these Maxwell terms do not affect dynamics drastically; the partition functions we present without Maxwell terms are expected to capture the same IR physics as the ones with the Maxwell terms and hence with clearer relation to the continuum field theories. Hence, in the remainder of the paper, we will omit Maxwell terms for the gauge fields. However, readers should note that this issue persists throughout any lattice dualities presented in this paper, as a manifestation of the same caveat in the continuum theories.

\subsection{Numerical Confirmation of the Implementation of ``Lagrange Multiplier''}
\label{sec:nuconf}
 
In this subsection, we present more detailed analysis of the lattice action $Z_{DW}$ in Eq.~\eqref{eq:doublewf}. We first present an analytical  argument that $Z_{DW}$ should admit a phase with long-ranged superconducting order that provides a Meissener effect for the gauge field $B-b$. When $b$ is promoted to being a fluctuating gauge field, this superconductivity higgses out $B-b$,  and $Z_{DW}$ effectively implements $\delta(B-b)$ in the deep IR. After the analytical argument, we present strong numerical evidence for the emergence of a superconducting phase in $Z_{DW}$. Once more, we emphasize $Z_{DW}$ \emph{does not} represent partition function of field theories that appear in dualities -- instead, superconductivity of $Z_{DW}[B,b,M,M]$ in certain parameter ranges of $M$ gives a strong evidence that $Z_{B}$ flows to a Wilson-Fisher fixed point in IR. Hence, our primary  interest in numerics is to see a gapped superconducting phase, not a critical phase. We reserve numerical studies of lattice actions corresponding to actual field theories in dualities and their critical points to future works. 

The analytical picture that a superconductor-insulator transition should be present in $Z_{DW}$ is the following. Using techniques similar to those used in the lattice construction of the bosonization duality \cite{Chen2018}, one can expand the exponentials on the lattice links  in the action $Z_{DW}$ as follows:
\begin{equation}
\begin{split}
& Z_{DW}[B,b;M_{1},M_{2}] = \int Da D\bar{\psi} D\psi D\bar{\chi} D\chi \, \left(\prod_{n} e^{M_{1}\bar{\psi}_{n} \psi_{n}} e^{M_{2}\bar{\chi}_{n} \chi_{n}}\right)\\
& \prod_{n\mu} \bigg[ 1 + \bar{\psi}_{n+\hat{\mu}}e^{i (a-b)_{n\mu}}\frac{\sigma^{\mu}-1}{2} \psi_{n} + \bar{\psi}_{n}e^{-i (a-b)_{n\mu}}\frac{-\sigma^{\mu}-1}{2} \psi_{n+\hat{\mu}} \\
& \hspace{1.1cm} +  \bar{\psi}_{n+\hat{\mu}}\frac{\sigma^{\mu}-1}{2} \psi_{n} \bar{\psi}_{n}\frac{-\sigma^{\mu}-1}{2} \psi_{n+\hat{\mu}} \bigg] \\[.1cm]
& \phantom{\prod_{n\mu}} \ \bigg[ 1 + \bar{\chi}_{n+\hat{\mu}}e^{-i (a-B)_{n\mu}}\frac{(\sigma^{\mu})^{*}-1}{2} \chi_{n} + \bar{\chi}_{n}e^{i (a-B)_{n\mu}}\frac{-(\sigma^{\mu})^{*}-1}{2} \chi_{n+\hat{\mu}} \\
& \hspace{1.1cm} + \bar{\chi}_{n+\hat{\mu}}\frac{(\sigma^{\mu})^{*}-1}{2} \chi_{n} \bar{\chi}_{n}\frac{-(\sigma^{\mu})^{*}-1}{2} \chi_{n+\hat{\mu}} \bigg].
\end{split}
\end{equation}
Here, $(\sigma^{\mu})^{*}$ denotes the complex conjugate (not Hermitian conjugate) of the Pauli matrices chosen for $Z_{W}$. Just as in the derivation of the lattice bosonization duality, here this expansion allows us to \textit{exactly} integrate out the gauge field. The idea is that, all terms in this expansion are either proportional to $e^{\pm i a_{n{\mu}}}$,  or have no dependence on $a_{n\mu}$. Those terms that are proportional to $e^{\pm i a_{n{\mu}}}$ \textit{vanish} upon integrating out the gauge field. This gives:
\begin{equation}
\label{eq:Z_aintegrated}
\begin{split}
& Z_{DW}[B,b;M_{1},U_{1},M_{2},U_{2}] = \int D\bar{\psi} D\psi D\bar{\chi} D\chi \, \prod_{n}\left( e^{M_{1}\bar{\psi}_{n} \psi_{n}} e^{M_{2}\bar{\chi}_{n} \chi_{n}}\right)\\
& \prod_{n\mu} \bigg( e^{i (B-b)_{n\mu}} \bar{\psi}_{n+\hat{\mu}}\frac{\sigma^{\mu}-1}{2} \psi_{n}\bar{\chi}_{n+\hat{\mu}}\frac{(\sigma^{\mu})^{*}-1}{2} \chi_{n} + e^{-i (B-b)_{n\mu}} \bar{\psi}_{n}\frac{-\sigma^{\mu}-1}{2} \psi_{n+\hat{\mu}}\bar{\chi}_{n}\frac{-(\sigma^{\mu})^{*}-1}{2} \chi_{n+\hat{\mu}} \\
& \ \ \ \ \ \ + \left[ 1 +  \bar{\psi}_{n+\hat{\mu}}\frac{\sigma^{\mu}-1}{2} \psi_{n} \bar{\psi}_{n}\frac{-\sigma^{\mu}-1}{2} \psi_{n+\hat{\mu}} \right] \left[ 1 +  \bar{\chi}_{n+\hat{\mu}}\frac{(\sigma^{\mu})^{*}-1}{2} \chi_{n} \bar{\chi}_{n}\frac{-(\sigma^{\mu})^{*}-1}{2} \chi_{n+\hat{\mu}}  \right] \bigg).
\end{split}
\end{equation}

Now, let us take a closer look at the second line of the above equation that contains the hopping terms that favor simultaneous hoppings of $\chi$ and $\psi$ fermions. For concreteness, we choose the following basis:
\begin{equation}
\begin{split}
&\sigma^{x} = \begin{pmatrix} 0 & 1 \\ 1 & 0 \end{pmatrix}, \, \sigma^{y} = \begin{pmatrix} 0 & i \\ -i & 0 \end{pmatrix}, \, \sigma^{z} = \begin{pmatrix} 1 & 0 \\ 0 & -1 \end{pmatrix}, \\[.2cm]
& \psi_{n} = \begin{pmatrix} \psi_{n,\uparrow} \\ \psi_{n,\downarrow} \end{pmatrix}, \,\bar{\psi}_{n} = \begin{pmatrix} \bar{\psi}_{n,\uparrow} \\ \bar{\psi}_{n,\downarrow} \end{pmatrix}, \, \chi_{n} = \begin{pmatrix} \chi_{n,\uparrow} \\ \chi_{n,\downarrow} \end{pmatrix} , \, \bar{\chi}_{n} = \begin{pmatrix} \bar{\chi}_{n,\uparrow} \\ \bar{\chi}_{n,\downarrow} \end{pmatrix}.
\end{split}
\end{equation}
In addition, it is convenient to define the following fermion bilinears (the motivation is that $(\sigma^{\mu} \pm 1)/2$ are projectors to the positive / negative eigenvectors of $\sigma^\mu$):
\begin{equation}
\begin{split}
&\phi_{n,0} = \frac{1}{\sqrt{2}}(\psi_{n,\uparrow}\chi_{n,\uparrow} + \psi_{n,\downarrow}\chi_{n,\downarrow}), \ \ \ \ \bar{\phi}_{n,0} = \frac{1}{\sqrt{2}}(\bar{\chi}_{n,\uparrow} \bar{\psi}_{n,\uparrow} + \bar{\chi}_{n,\downarrow} \bar{\psi}_{n,\downarrow} )  \\
&\phi_{n,x} = \frac{1}{\sqrt{2}}(\psi_{n,\uparrow}\chi_{n,\downarrow} + \psi_{n,\downarrow}\chi_{n,\uparrow}), \ \ \ \ \bar{\phi}_{n,x} = \frac{1}{\sqrt{2}}(\bar{\chi}_{n,\downarrow} \bar{\psi}_{n,\uparrow} + \bar{\chi}_{n,\uparrow} \bar{\psi}_{n,\downarrow})  \\
&\phi_{n,y} = \frac{-i}{\sqrt{2}}(\psi_{n,\uparrow}\chi_{n,\downarrow} - \psi_{n,\downarrow}\chi_{n,\uparrow}), \ \ \ \ \bar{\phi}_{n,y} = \frac{i}{\sqrt{2}}(\bar{\chi}_{n,\downarrow} \bar{\psi}_{n,\uparrow} - \bar{\chi}_{n,\uparrow} \bar{\psi}_{n,\downarrow} ) \\
&\phi_{n,z} = \frac{1}{\sqrt{2}}(\psi_{n,\uparrow}\chi_{n,\uparrow} - \psi_{n,\downarrow}\chi_{n,\downarrow}), \ \ \ \ \bar{\phi}_{n,z} = \frac{1}{\sqrt{2}}(\bar{\chi}_{n,\uparrow} \bar{\psi}_{n,\uparrow} - \bar{\chi}_{n,\downarrow} \bar{\psi}_{n,\downarrow}).
\end{split}
\end{equation}
These fermion bilinears can be understood as hard-core bosons formed out of two fermions, and they enable us to write down Eq.~\eqref{eq:Z_aintegrated} in a more compact way:
\begin{equation}
\begin{split}
& Z_{DW}[B,b;M_{1},U_{1},M_{2},U_{2}] = \int D\bar{\psi} D\psi D\bar{\chi} D\chi \, \prod_{n}\left( e^{M_{1}\bar{\psi}_{n} \psi_{n}} e^{M_{2}\bar{\chi}_{n} \chi_{n}}\right)\\
& \prod_{n\mu} \bigg( e^{i (B-b)_{n\mu}} (\bar{\phi}_{n+\hat{\mu},0} - \bar{\phi}_{n+\hat{\mu},\mu})(\phi_{n,0} - \phi_{n,\mu}) + e^{-i(B-b)_{n\mu}} (\bar{\phi}_{n,0} + \bar{\phi}_{n,\mu})(\phi_{n+\hat{\mu},0} + \phi_{n+\hat{\mu},\mu}) \\
& \ \ \ \ \ \ + \left[ 1 +  \bar{\psi}_{n+\hat{\mu}}\frac{\sigma^{\mu}-1}{2} \psi_{n} \bar{\psi}_{n}\frac{-\sigma^{\mu}-1}{2} \psi_{n+\hat{\mu}} \right] \left[ 1 +  \bar{\chi}_{n+\hat{\mu}}\frac{(\sigma^{\mu})^{*}-1}{2} \chi_{n} \bar{\chi}_{n}\frac{-(\sigma^{\mu})^{*}-1}{2} \chi_{n+\hat{\mu}}  \right] \bigg).
\end{split}
\end{equation}
Let's focus on the hardcore bosons in the second line. Note that hardcore bosons of the form $\phi_{n,\mu}$ can only hop across links along the $\hat{\mu}$ direction; they effectively ``live" in one dimension, and in the absence of any coupling between them, they cannot condense in the thermodynamic limit due to the Mermin-Wagner theorem\footnote{Of course, couplings between them are not forbidden by symmetry and will therefore be generated.  To the extent that such terms are small, the transition temperature at which  long range order develops will be parametrically small.}.   By contrast, the $\phi_{n, 0}$ boson hops  along all three lattice directions, and may condense under appropriate choice of microscopic parameters.  Thus, we may neglect $\phi_{n, \mu}$, integrate out the fermions, and study the effective theory for $\phi_{n,0}$.  Away from critical points, the effective action for $\phi_{n, 0}$ will be local and takes the usual Ginzburg-Landau-Wilson form.  When the quadratic term is negative, the $\phi_{n,0}$ boson will acquire long range order, and this corresponds to a superconducting phase.  Since $\phi_{n,0}$ is charged under the gauge field $B-b$, it expels precisely this combination of gauge fields via the Meissner effect in the superconducting phase.  In the present argument, the role of the fluctuating gauge field $a$  is to  ``glue" the  two fermions $\psi, \chi$, and when this composite field condenses, a superconducting phase results.  

\begin{figure}
\centering
\begin{minipage}{0.49\linewidth}
\includegraphics[width = 1.0\linewidth]{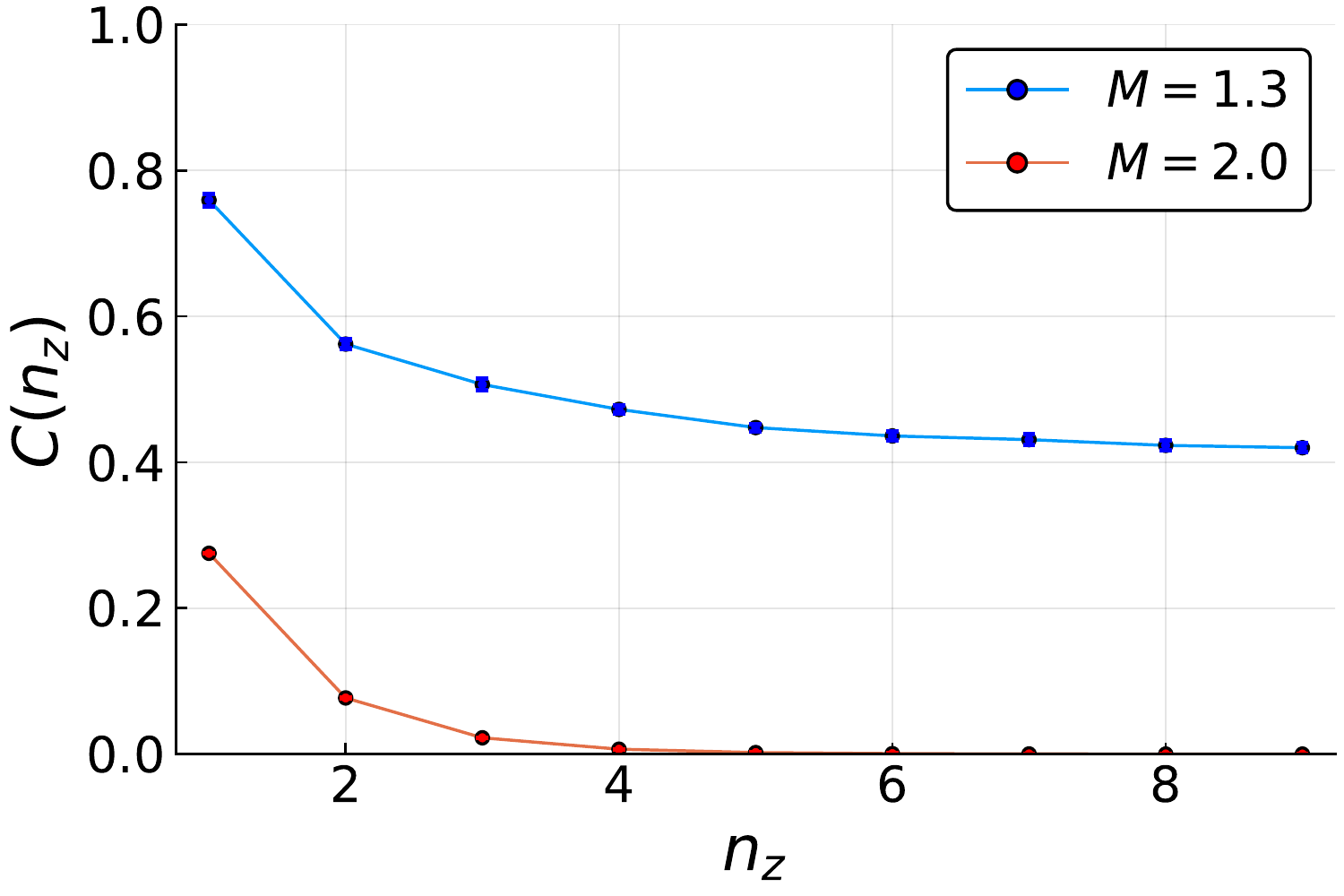}
\end{minipage}
\begin{minipage}{0.49\linewidth}
\includegraphics[width = 1.0\linewidth]{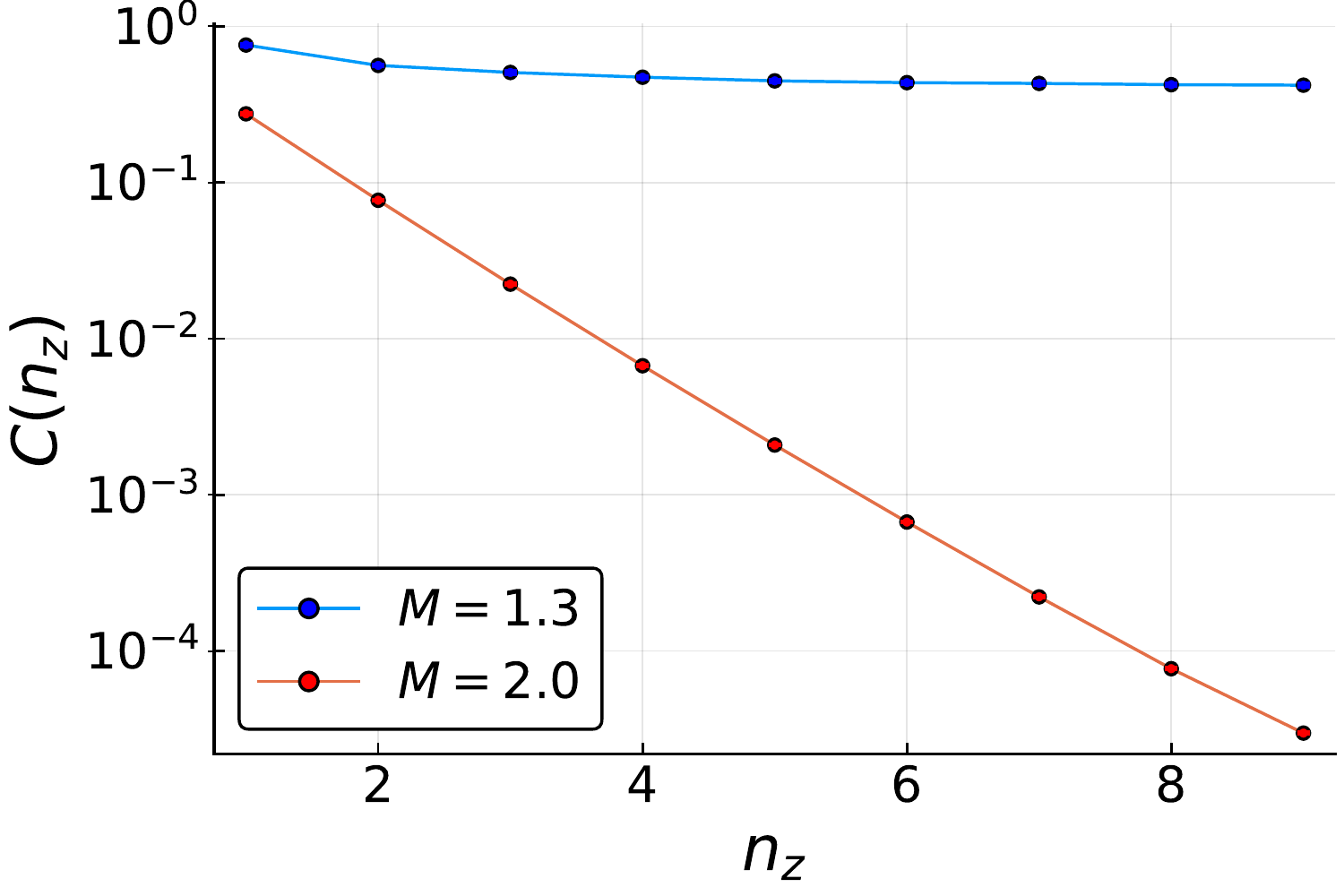}
\end{minipage}
\caption{Plots of a correlation function $C(n_{z}) = \left\langle \bar{\phi}_{n_1,0} \phi_{n_2,0}\right\rangle$ for separation $n_2-n_1= (0,0, n_z)$ versus $n_{z}$, at $M=1.3$ (blue) and $M=2.0$ (red), in linear-linear scale (left panel) and log-linear scale (right panel). Superconducting long-range order is present at $M=1.3$ and absent at $M=2$.}
\label{fig:corgraph1}
\end{figure}

To demonstrate the validity of the above picture explicitly, we sample $Z_{DW}$ directly in Eq.~\eqref{eq:doublewf} in a determinant quantum Monte Carlo (DQMC) simulation.  When $M_{1} = M_{2} =M $, $B = b = 0$, one can compute the above correlator from the determinant-Monte-Carlo like algorithm\footnote{The requirement $M_1 = M_2$ is merely to avoid the fermion sign problem.  Recall that for energy scales $\mu \ll M_1, M_2$, the precise values of these masses are unimportant.  Consequently, we set them equal in order to avoid the fermion sign problem.}. Note that while the theory is strongly interacting due to the presence of the fluctuating gauge field $a$, the fermions can exactly be integrated out on the lattice for each configuration of $a$.  The resulting fermion determinants come in complex-conjugate pairs, and there is no sign problem -- as can be seen from the fact that the integrand of Eq.~\eqref{eq:doublewf} is manifestly positive.  The details of numerical implementations are explained in the appendix.
Based on the analytic picture above, we identify the following correlator that diagnoses off-diagonal long-range order that signals the emergence of superconductivity:
\begin{equation}
\label{eq:corcondense}
\left\langle \bar{\phi}_{n_{1},0} \: \phi_{n_{2},0}\right\rangle.
\end{equation}

We present the simulation results done on the system with size $8 \times 8 \times 20$, with periodic boundary conditions on all directions. We chose one direction longer than the others to see long distance behavior of correlators more clearly; since the system we are investigating is isotropic, it is natural to expect that the behavior along one direction persists in other directions as well. 
 
We see a clear signature of off-diagonal long-range orders, as shown by the blue curves in Fig.~\ref{fig:corgraph1}, for low values of $M$: The distance-correlator plot in linear-linear scale shows that at the long distance the correlator reaches some constant, signaling emergence of long-range order. Repeating the same analysis on $1<M<4$ we found the transition point $M \approx 1.8$ where the correlation function behavior sharply transitions into that of an exponentially decaying function: The red curves in Fig.~\ref{fig:corgraph1}, especially in linear-log scale, show rapid exponential decay of correlator with respect to distance. In Fig.~\ref{fig:corgraph1} and all other graphs presenting the results of Monte-Carlo simulations, error bars are explicitly included. However, in case of determining emergence of superconductivity from the correlator Eq.~\eqref{eq:corcondense}, correlators can be computed with very small errors, often the size of errors on the graphs being smaller than thickness of two bars that represent the range of the error.
 
\begin{figure}
\centering
\includegraphics[width = 0.5\linewidth]{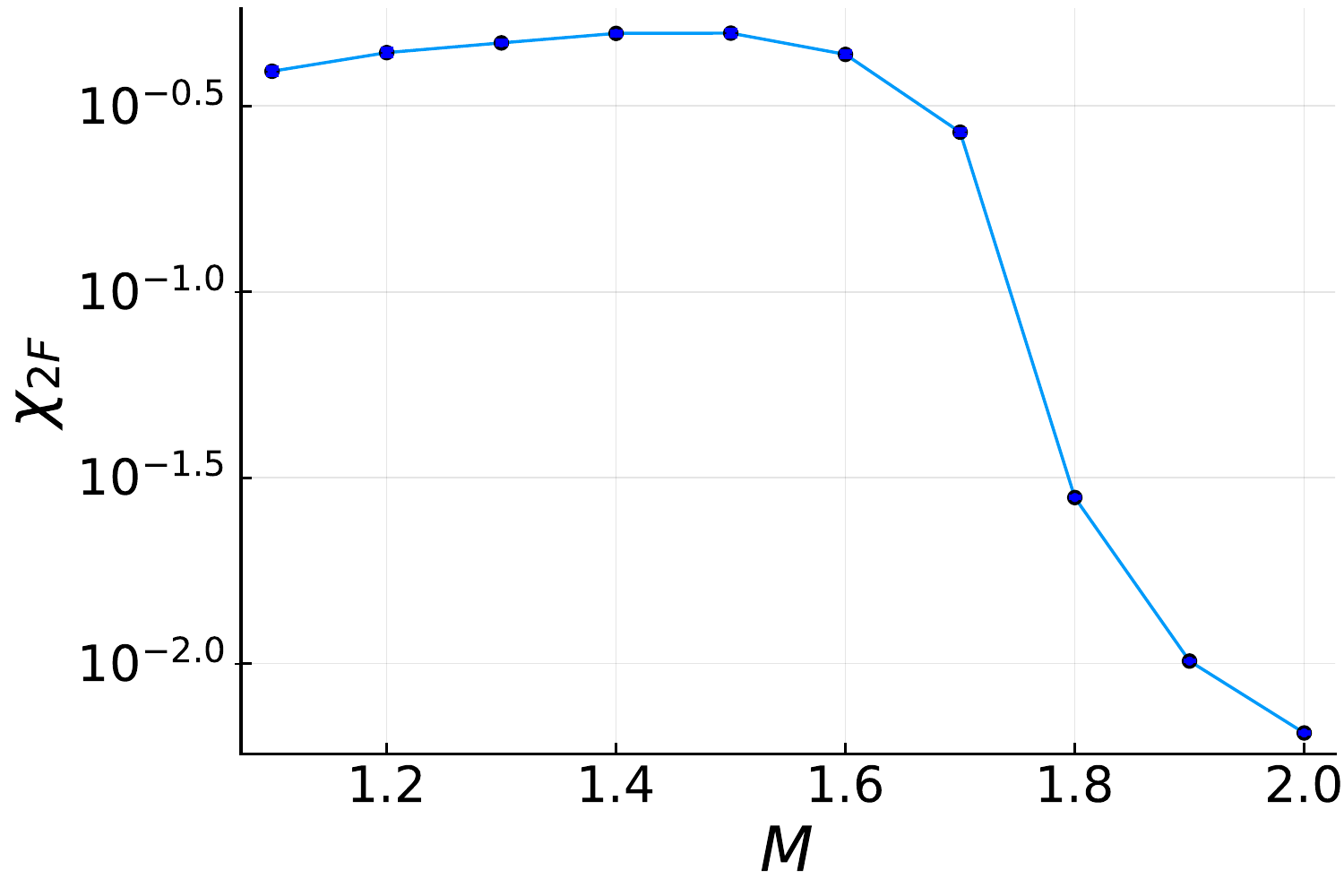}
\caption{Plot of $\chi_{2F}$ versus $M$ for $M = 1.1 - 2.0$}
\label{fig:suspre}
\end{figure}

 Another interesting quantity that probes long-range order and signals sharp transition at $M\approx 1.8$ is $\chi_{2F}$ (in continuum, this can be interpreted as a renormalized pair susceptibility), defined as:
\begin{equation}
 \chi_{2F} = \frac{1}{V^2}\sum_{n_{1},n_{2}} \left\langle \bar{\phi}_{n_{1},0} \: \phi_{n_{2},0}\right\rangle.
\end{equation}
The plot of $\chi_{2F}$ for different mass values in Fig.~\ref{fig:suspre} clearly shows that this number sharply drops at $M\approx 1.8$, indicating loss of long-range order at this $M$ value.
 
We present this Monte-Carlo result as a clear evidence that the theory defined by $Z_{DW}$ exhibits a Higgsing of the gauge field $B-b$ (at least for small $B-b$) in the IR. That the phase transition occurs at $M\approx 1.8$ rather than $M\approx 3$ is because interaction in general changes UV parameters; it will take different values if we include a Maxwell term for $a$, which we will explore in future works.


\subsection{On the Boson-Vortex duality}
\label{sec:comment_on_BV}

So far, we have observed that including additional heavy Wilson fermions to the  euclidean lattice construction of the 3D bosonization duality allows one to realize other dualities in the $U(1)$ duality web as exact partition function mappings. However, we did not present how similar methods would  realize the boson-vortex duality. Here, we illustrate the difficulty of realizing the  boson-vortex duality as  an exact partition function mapping; 
nevertheless, we provide here some numerical evidence for the claim that topological terms involving heavy Wilson fermions can realize this duality.  

Recall that in the continuum derivation of the boson-vortex duality, we used something that we have not otherwise used in the previous subsections: the two equivalent expression of \eqref{eq:Tu1f}, which relies on
\begin{equation}
-\mathcal{L}_{\rm Dirac}^{+, m} [A] \ = \ -\mathcal{L}_{\rm Dirac}^{-, m} [A] + \frac{i}{4\pi} AdA.
\end{equation}
While this equality is innocent in the IR, such an identification cannot be carried over to the UV. For instance, if we implement these IR Lagrangians by Wilson fermions on the lattice, as we did above, then the left-hand-side is implemented by $Z_W[A; 3+m, 0]$, while the right-hand-side by $Z^\ast_W[A; 3-m, 0] Z_W[A; M_h]$ for some $1<M_h<3$. The two implementations certainly are different in the UV. As the continuum ``derivation'' of the boson-vortex duality from the bosonization duality relied on the identification above, this procedure does not yield an exact mapping on the lattice for the boson-vortex duality.

Below, we present a lattice implementation of the Abelian Higgs model using the elements we have introduced, though this implementation does not have exact mapping relations to the previous lattice bosonization duality via any lattice $\mathcal{S}$ and $\mathcal{T}$ transformation. Our strategy is to construct the BF Lagrangian  out of level 1 Chern-Simons terms as the following\footnote{The reason we have the include the seemly useless $A$ is that, the connection in a level 1 Chern-Simons must be spin-$c$ instead of $U(1)$, because a level 1 Chern-Simons is naturally related to a massive Dirac fermion, which couples to a spin-$c$ connection. In fact, all topological gauge terms consistent with the spin-$c$ property may be built out of linear combination of level 1 Chern-Simons terms.}:
\begin{equation}
\frac{i}{2\pi} bdB = \frac{i}{4\pi} (A+b+B)d(A+b+B) - \frac{i}{4\pi} (A+b)d(A+b) - \frac{i}{4\pi} (A+B)d(A+B) + \frac{i}{4\pi} AdA.
\end{equation}

Motivated by this expression as well as the numerical result in Sec.~\ref{sec:nuconf}, consider the following partition function:
\begin{equation}
\label{eq:abhiggs2}
Z_{\rm{AbHiggs}}[A,B;T,M] = \int Db \ Z_{XY}[b;T] \ Z_{BF}[A,b,B;M]
\end{equation}
where $Z_{BF}[A,b,B;M]$, the term that implements $\frac{i}{2\pi} b dB$ via heavy fermions, is defined as\footnote{In $Z_{BF}[A,b,B;M]$, the spin-$c$ connection $A$ is included for heavy fermions to satisfy the spin-charge relation -- its role is purely auxiliary and may be assumed to be zero for the analysis.}:
\begin{equation}
\label{eq:BFheavy}
\begin{split}
Z_{BF}[A,b,B;M] = & Z_{W}[A+b+B;M,0]Z_{W}^{*}[A+b;M,0]\\
& Z_{W}^{*}[A+B;M,0]Z_{W}[A;M,0]
\end{split}
\end{equation}
Observe that only the first two heavy fermions of $Z_{BF}[A,b,B;M]$ contribute to the dynamics, since the latter two are not coupled to dynamical gauge fields. We saw in Sec.~\ref{sec:nuconf} that by tuning $M$ to a proper value and sending $T\rightarrow \infty$, one can enter the phase in which a gauge-invariant fermion bilinear condenses. Meanwhile, when $T \rightarrow 0$, $Z_{XY}[b;T]$ Higgs the gauge field. Since the gauge fluctuation plays a key role in condensation of the gauge-invariant fermion bilinear, Higgsing out the gauge field will ``decondense' the composite objects of two fermions. Thus, the partition function faithfully captures a theory in which a bosonic object condenses (stays trivially gapped) in high temperature (low temperature). Interpreting the fermion bilinear as a monopole operator is consistent with both the phase diagram of the Abelian Higgs model and the recurrent motif of the paper in which lattice heavy fermion operators are often understood as monopole operators in continuum.
 
When Peskin \cite{Peskin1977} first suggested the boson-vortex duality, he demonstrated it with a lattice model. The advantage in our lattice implementation, however, is that it explicitly preserves $U(1)$ nature of the gauge fields -- i.e. the gauge field cannot be arbitrarily rescaled, and the level must be quantized. The monopole operators are manifestly available and their condensation is explicit.
 
Based on above picture, we conjecture that $Z_{\rm{AbHiggs}}[B;T,M]$ and $Z_{\rm{XY}}[B;T']$ are IR dual, with the gauge-invariant fermion bilinear in $Z_{\rm{AbHiggs}}[B;T,M]$ corresponding to the complex boson $e^{\pm i\theta}$. The theory $Z_{\rm{AbHiggs}}$ at $B=0$ can be simulated with sign-free Monte-Carlo as well, since all one has to do from the simulation presented in Sec.~\ref{sec:nuconf} is to add ``weight' of the each gauge field configuration coming from $XY$ part of the action. One may able to extract critical exponents of $Z_{\rm{AbHiggs}}[B;T,M]$ and compare them with $XY$ model critical exponents. We reserve verifying this claim numerically to the future work.

\subsection{Implementing generic $\mathcal{T}$ and $\mathcal{S}$ transformation on the Euclidean lattice}
\label{sec:more_S_T}

First, let us explore the general structure of continuum quantum field theories in the duality web. One can generically write any element $\tau \in \text{PSL}(2, \mathbb{Z})$ as:
\begin{equation}
\label{eq:gtf}
\tau = \left( \prod_{i=1}^{n}\mathcal{T}^{m_{i}}\mathcal{S} \right) \mathcal{T}^{m_{n+1}}.
\end{equation}
Here, $m_{i}$ can take any integer value. Given a Lagrangian $-\mathcal{L}[A]$ with the background $U(1)$ gauge field in the Eulicdean signature, $\tau$, as defined in Eq.~\eqref{eq:gtf} corresponds to the following modification of the Lagrangian:
\begin{equation}
\label{eq:gtflag}
\tau : -\mathcal{L}[A] \rightarrow -\mathcal{L}[a_{1}] + \frac{i}{4\pi} K_{IJ} a_{I} d a_{J}.
\end{equation}
Here, $I$ and $J$ are indices that run from $1$ to $n+1$. We define $a_{I} = (a_{1}, a_{2}, \cdots, a_{n}, A)$; $a_{1}$, $a_{2}$, $\cdots$, $a_{n}$ corresponds to the fluctuating gauge field introduced by $\mathcal{S}$ transformation, and $A = a_{n+1}$ corresponds to the background gauge field. $K_{IJ}$ is a symmetric matrix with integer entry, defined as the following:
\begin{equation}
\label{eq:formkij}
K_{IJ} = \begin{pmatrix}
m_{1} & 1 & 0 & \cdots & 0 &0 \\
1 & m_{2} & 1 & \cdots & 0 &0 \\
0 & 1 & m_{3} & \cdots & 0 &0 \\
\vdots & \vdots & \vdots & \ddots & \vdots &\vdots \\
0 & 0 & 0 & \cdots & m_{n} &1 \\
0 & 0 & 0 & \cdots & 1 & m_{n+1} 
\end{pmatrix}.
\end{equation}
Diagonal elements $K_{II} = m_{I}$ corresponds to $\mathcal{T}^{m_{I}}$ transformation in Eq.~\eqref{eq:gtf}, and off-diagonal elements correspond to BF terms introduced by $\mathcal{S}$ transformations in Eq.~\eqref{eq:gtf}. 
 

 The following two considerations allow one to write down Eq.~\eqref{eq:gtf} in a slightly different form:
\begin{itemize}
\item one has a freedom to add any integer multiples of $A$ into the fluctuating gauge field to get an equivalent expression, i.e. 
\begin{equation}
a_{i} \rightarrow \tilde{a}_{i}  = a_{i} + l_{i} A
\end{equation}
for $i =1 ,\cdots , n$ give equivalent expression since this is tantamount to adding constant to the integration variable.
\item The quantum gauge fields $a_{1}$, $a_{2}$, $\cdots$, $a_{n}$ can be ``mixed' by SL$(n,\mathbb{Z})$ transformations. For a $n \times n$ matrix $S \in \text{SL}(n,\mathbb{Z})$ The change of the variable
\begin{equation}
\begin{pmatrix}
a_{1} \\
a_{2} \\
\vdots \\
a_{n}
\end{pmatrix} \rightarrow \begin{pmatrix}
\tilde{a}_{1} \\
\tilde{a}_{2} \\
\vdots \\
\tilde{a}_{n}
\end{pmatrix} = S \begin{pmatrix}
a_{1} \\
a_{2} \\
\vdots \\
a_{n}
\end{pmatrix}
\end{equation}
preserves integration measure and compact nature of $U(1)$ gauge fields. Thus, rewriting Lagrangian on the right hand side of Eq.~\eqref{eq:gtflag} in terms of $\tilde{a}_{1}$, $\tilde{a}_{2}$, $\cdots$, $\tilde{a}_{n}$ is a valid variable change.
\end{itemize}

 Above two variable changes can be accomodated by considering the variable change given by a matrix $M \in \text{SL}(n+1, \mathbb{Z})$ with some constraint:
\begin{equation}
a_{I} \rightarrow \tilde{a_{I}} = M_{IJ} a_{J}, \quad M_{n+1,I} = \delta_{n+1,I}.
\end{equation}
 The constraint on the last row of $M$ is present to make sure that the definition of the background gauge field is not modified. Defining $(M^{-1})^{T} K M^{-1} = \tilde{K}$, Eq.~\eqref{eq:gtflag} can be equivalently rewritten as:
\begin{equation}
\label{eq:gtflag2}
\tau : -\mathcal{L}[A] \rightarrow -\mathcal{L}[(M^{-1})_{1I}\tilde{a}_{I}] + \frac{i}{4\pi} \tilde{K}_{IJ} \tilde{a}_{I} d \tilde{a}_{J}.
\end{equation}
Note that $\tilde{K}_{IJ}$ is still real symmetric matrix but may not be a band diagonal matrix as the original $K$-matrix in Eq.~\eqref{eq:formkij} is.

The $K$-matrix multi-component Chern-Simons term Eq.~\eqref{eq:gtflag2} primarily consist of two terms: The BF term that couple two different gauge fields and the diagonal Chern-Simons term that simply reduce to the Chern-Simons term for single gauge field. The important point is that, \textit{all $K$-matrix elements have corresponding lattice heavy fermion representations}. For example, we know that under the proper choice of the parameter $M$ (for later convenience, we set the four-fermi counterterm $U=0$ for implementing P$SL(2,\mathbb{Z})$ transformation on the lattice),
\begin{equation}
Z_{W}[\pm A;M,U=0] \xrightarrow[]{\text{IR limit}} e^{\frac{i}{4\pi}\int A dA }
\end{equation}
Note that the sign of the charge of the heavy fermion with respect to the gauge coupling $A$ do not affect the coefficient of IR Chern-Simons term. To generate level $-1$ Chern-Simons term, consider the action $Z_{W}]^{*}[\pm A;M,U=0$. This action is almost identical to $Z_{W}$, except pauli matrices $\sigma^{\mu}$ are substituted with its complex conjugate version (Note that we are not taking Hermitian conjugation, under which Pauli matrices remain same) and $A$ is substituted with $-A$. As the superscript suggests, the action $Z_{W}^{*}[\pm A;M,U=0]$ is precisely complex conjugate of $Z_{W}[\pm A;M,U=0]$. Hence,
\begin{equation}
Z_{W}^{*}[\pm A;M,U=0] \xrightarrow[]{\text{IR limit}} e^{\frac{-i}{4\pi}\int A dA }.
\end{equation}
A Chern-Simons term with higher levels can be simply generated by placing multiple heavy fermions with the same gauge field coupling.

 Now, let us consider the heavy fermion representation of the BF coupling. Note that
\begin{equation}
\begin{split}
\int \frac{i}{2\pi} A_{1} d A_{2} &= \int \frac{i}{4 \pi} (A_{1}+A_{2}) d (A_{1}+A_{2}) - \frac{i}{4\pi} A_{1} d A_{1} - \frac{i}{4\pi} A_{2} d A_{2} \\
&= \int - \frac{i}{4 \pi} (A_{1}-A_{2}) d (A_{1}-A_{2}) + \frac{i}{4\pi} A_{1} d A_{1} + \frac{i}{4\pi} A_{2} d A_{2}.
\end{split}
\end{equation}
Hence, we expect following heavy fermion actions to realize BF coupling in IR:
\begin{equation}
\begin{split}
Z_{W}[\pm (A_{1}+A_{2});M,0]Z_{W}^{*}[\pm A_{1};M,0]Z_{W}^{*}[\pm A_{2};M,0 ] & \xrightarrow[]{\text{IR limit}} e^{\frac{i}{2\pi}\int A_{1}d A_{2} } \\
Z_{W}^{*}[\pm (A_{1}A_{2});M,0]Z_{W}[\pm A_{1};M,0]Z_{W}[\pm A_{2};M,0 ] & \xrightarrow[]{\text{IR limit}} e^{\frac{i}{2\pi}\int A_{1}d A_{2} }.
\end{split}
\end{equation}
Similar to the case of the lattice heavy fermion Chern-Simons action, ``Complex-conjugated version" of the fermion action on the left side of the above equations can generate BF coupling with opposite sign. Also, BF-coupling with integer coefficients other than $\pm 1$ can be generated by multiplying multiple lattice actions, each of which generate unit-level BF couplings.

We conclude this subsection with two remarks. First, there are a number of different ways to implement general P$SL(2,\mathbb{Z})$ transformation on the lattice, primarily due to two factors: There is a freedom of changing variable with respect to the restricted $\text{SL}(n+1,\mathbb{Z})$ transformations to obtain different $K$ matrices, and there are multiple ways of implementing the same BF couplings/Chern-Simons terms with lattice heavy fermions. We argue that these different lattice actions all flow to the same IR fixed point, yet we will see that some non-trivial properties of dualities in the web is more manifest in a certain lattice action. Second, the ``free-fermion picture" tells us that $M$ should be tuned to be $1<|M|<3$ to implement a Chern-Simons term with level $\pm 1$, However, in actual implementations, gauge fields coupled to the heavy fermion is quantum gauge fields, and their fluctuations generally renormalize the range of $M$ in which the lattice action is expected to flow to the IR fixed point we want.

\section{A Microscopic Model for $N_f=2$ Self-Duality}
\label{sec:nf2}

In this section, we focus on the self-dual $N_f=2$ three-dimensional QED \cite{Karch2016, Xu2015, Benini2017}. This duality has found applications in condensed matter physics, most interestingly as an effective description of surface states of $(3+1)D$ bosonic topological insulators \cite{Xu2015, Senthil2012} and deconfined quantum criticality \cite{Wang2017}. This duality is remarkable for it being self-dual, as we will see soon, and has an emergent $SU(2)\times SU(2)$ global symmetry at criticality, as we will discuss in later subsections. The goal of this section is to explicitly construct the operator maps of the duality, and also provide a picture for how the $SU(2)\times SU(2)$ symmetry emerges at the criticality.
 
Consider the duality between the following three theories:
\begin{equation}
\label{eq:dualnf2}
\begin{split}
-\mathcal{L}_{N_{f}=2}[B,B'] =& - \mathcal{L}_{\rm Dirac \psi_1}^{-, -m_1}[a+B]  - \mathcal{L}_{\rm Dirac \psi_2}^{+, m_2}[a-B] + \frac{i}{2 \pi} a d (B'+B) - \frac{i}{4 \pi} (B+B') d (B+B')  \\
\updownarrow & \\
-\mathcal{L}_{N_{b}=2}[B,B'] =& -\mathcal{L}_{\rm WF \phi_1}^{r_1}[b+B] -\mathcal{L}_{\rm WF \phi_2}^{r_2}[b+B'] + \frac{i}{2\pi}b d(B+B') \\
 \updownarrow & \\  
-\tilde{\mathcal{L}}_{N_{f}=2}[B,B'] =& - \mathcal{L}_{\rm Dirac \chi_1}^{+, m_1}[\tilde{a}-B']  - \mathcal{L}_{\rm Dirac \chi_2}^{-, -m_2}[\tilde{a}+B'] + \frac{i}{2 \pi} \tilde{a} d (B'+B) - \frac{i}{4 \pi} (B+B') d (B+B')
\end{split}
\end{equation}
where $\sgn(r_1)=\sgn(m_1), \sgn(r_2)=\sgn(m_2)$. Note that $-\mathcal{L}_{N_{f}=2}[B,B']$ and $-\tilde{\mathcal{L}}_{N_{f}=2}[B,B']$ are almost identical when $m_1=m_2$, except the roles of $B$ and $B'$ have been exchanged. Thus, the fermion theory presented in $-\mathcal{L}_{N_{f}=2}[B,B']$ and  $-\tilde{\mathcal{L}}_{N_{f}=2}[B,B']$ is referred as being self-dual. 

Before we discuss it on the lattice, let us review how this duality can be obtained from the duality web that we discussed in the previous section. Let us start from two copies of the Wilson-Fisher bosons:
\begin{equation}
\begin{split}
-\mathcal{L}_{{\rm DWF}}[X,Y] =& -\mathcal{L}_{\rm WF \phi_1}^{r_1}[X] -\mathcal{L}_{\rm WF \phi_2}^{r_2}[Y]
\end{split}
\end{equation}
One can consider applying the following series of modifications:
\begin{enumerate}
\item Substitute $X$ with $B_{f} + B$ and $Y$ with $B_{f}+B'$. Note that $X$ and $Y$ are $U(1)$ background gauge fields, so such substitution is benign. Also, such substitution respects compactness of gauge fields (though the inverse of this substitution does not, because this substitution changes the number of gauge fields).
\item Attach the BF term $\frac{i}{2\pi}B_{f} d(B+B')$.
\item Promote $B_{f}$ to the fluctuating gauge field $b$.
\end{enumerate}

The above procedure modifies $-\mathcal{L}_{{\rm DWF}}[X,Y]$ to $-\mathcal{L}_{N_{b}=2}[B,B']$. Meanwhile, we know through the duality Eq.~\eqref{eq:fbf} (i.e. Eq.~\eqref{eq:fbdual1} and its time-reversed version) that $-\mathcal{L}_{{\rm DWF}}[X,Y]$ should be dual to the following fermionic theories:
\begin{equation}
\begin{split}
 -\mathcal{L}_{DF}[X,Y] & = - \mathcal{L}_{\rm Dirac \psi_1}^{-, -m_1}[a_1] + \frac{i}{4\pi} (a_{1}-X) d (a_{1}-X) - \mathcal{L}_{\rm Dirac \psi_2}^{+, m_2}[a_2]  - \frac{i}{4\pi} (a_{2}-Y) d (a_{2}-Y) \\[.1cm]
 -\tilde{\mathcal{L}}_{DF}[X,Y] & = - \mathcal{L}_{\rm Dirac \chi_1}^{+, m_1}[\tilde{a}_1] - \frac{i}{4\pi} (\tilde{a}_{1}-X) d (\tilde{a}_{1}-X) - \mathcal{L}_{\rm Dirac \chi_2}^{-, -m_2}[\tilde{a}_2] + \frac{i}{4\pi} (\tilde{a}_{2}-Y) d (\tilde{a}_{2}-Y) \\
\end{split}
\end{equation}
Repeating the procedure we applied to $\mathcal{L}_{\rm DWF}$, we find $ -\mathcal{L}_{DF}[X,Y] $ and $ -\tilde{\mathcal{L}}_{DF}[X,Y]$ become $-\mathcal{L}_{N_{f}=2}[B,B']$ and $-\tilde{\mathcal{L}}_{N_{f}=2}[B,B']$ respectively. The key step in the manipulation is that after modification, the fluctuating gauge field $b$ only appear in the following BF terms:
\begin{equation}
\begin{split}
\frac{i}{2\pi} b d (a_{2}-a_{1} + 2B) \qquad \text{in $-\mathcal{L}_{N_{f}=2}[B,B']$} \\[.1cm]
\frac{i}{2\pi} b d (\tilde{a}_{1}-\tilde{a}_{2} + 2B') \qquad \text{in $-\tilde{\mathcal{L}}_{N_{f}=2}[B,B']$}
\end{split}
\end{equation}
In both cases, one can explicitly integrate out $b$ to implement $\delta(a_{2}-a_{1} + 2B)$ and $\delta(\tilde{a}_{1}-\tilde{a}_{2} + 2B')$. This allows the substitution $a_{1} \rightarrow a + B$, $a_{2} \rightarrow a - B$, $\tilde{a}_{1} \rightarrow \tilde{a} - B'$ and $\tilde{a}_{2} \rightarrow \tilde{a} + B'$. In this substitution, compactness of the gauge field is preserved, given that $a_{1,2}, \tilde{a}_{1,2}$ satisfy the delta function constraints from $b$.

\subsection{Lattice Realization of $N_f=2$ Self-Duality}
 We will invoke the above procedure on the lattice partition functions to obtain the lattice Euclidean version of the duality Eq.~\eqref{eq:dualnf2}. Using the exact lattice mapping \eqref{eq:fbduallat} and its time reversal version (referring to Eq.~\eqref{eq:timeRdualF}), the following three partition functions are exactly equivalent:
\begin{equation}
\begin{split}
&Z_{DWF}[X,Y;T_{1,2},M] = \int Da_{1} Da_{2} Db_{1} Db_{2} \, Z_{\rm{XY}}[b_{1};T_1]Z_{W}^{*}[a_{1}-b_{1};M,0]Z_{W}[a_{1}-X; M, 0] \\
& \hspace{7.7cm} Z_{\rm{XY}}[b_{2};T_2]Z_{W}[a_{2}-b_{2};M,0]Z_{W}^{*}[a_{2}-Y; M, 0] \\
&Z_{DF}[X,Y;M'_{1,2},U'_{1,2},M] = \int Da_{1} Da_{2} \, Z_{W}^{*}[a_{1};M'_1,U'_1]Z_{W}[a_{1}-X; M, 0] \\
& \hspace{6.9cm} Z_{W}[a_{2};M'_2,U'_2]Z_{W}^{*}[a_{2}-Y; M, 0] \\
&\tilde{Z}_{DF}[X,Y;M'_{1,2},U'_{1,2},M] = \int D\tilde{a}_{1} D\tilde{a}_{2} \, Z_{W}[-\tilde{a}_{1};M'_1,U'_1]Z_{W}^{*}[-\tilde{a}_{1}+X; M, 0] \\
& \hspace{6.9cm} Z_{W}^{*}[-\tilde{a}_{2};M'_2,U'_2]Z_{W}[-\tilde{a}_{2}+Y; M, 0]
\end{split}
\end{equation}
We remind that $Z_{DF}$ is obtained from $Z_{DWF}$ by the exactly mapping $\int Db_{1} Z_{\rm{XY}}[b_{1}]Z_{W}^{*}[a_{1}-b_{1}]$ to $Z_{W}^{*}[a_{1}]$ and $\int Db_{2} Z_{\rm{XY}}[b_{2}]Z_{W}[a_{2}-b_{2}]$ to $Z_{W}[a_{2}]$; on the other hand, to obtain $\tilde{Z}_{DF}$, we first replace $a_1=-\tilde{a}_1+b_1+X$, $a_2=-\tilde{a}_2+b_2+Y$, and then use the same exact mappings that integrate out $b_{1,2}$. The Wilson fermions with mass $M'$ are those that correspond the light Dirac fermions in $\mathcal{L}_{DF}$ and $\tilde{\mathcal{L}}_{DF}$ in the continuum, while those with mass $M$ corresponds to the monopole operators in the Chern-Simons terms. Note that in the duality between $Z_{DF}$ and $\tilde{Z}_{DF}$, the roles of light fermions and monopole operators (heavy fermions) are switched; moreover, the time reversal operation leaving $Z_{DWF}$ invariant maps $Z_{DF}$ and $\tilde{Z}_{DF}$ to each other.

Now, one can do a substitution $X \rightarrow B_{f} +B$, $Y \rightarrow B_{f}+B'$, and include a factor $Z_{BF}[A,B+B',B_{f};M]$ (see Eq.~\eqref{eq:BFheavy} for definition) that plays the role of the BF term. (Recall that we included the background spin-$c$ connection $A$ to make heavy fermions satisfy spin-charge relations; its role is purely auxiliary. We just set $A=0$.) As a final step, turn $B_{f}$ to the fluctuating gauge field $b$. The string of procedures we described is simply a lattice version of the manipulation we described earlier to obtain $N_{f}=2$ self-duality from two copies of the boson-fermion duality described. However, we emphasize that the fact that continuum field theory dualities in this paper are IR dualities implies that promoting the gauge field $b$ from the classical one to the quantum one is subtle, the aforementioned procedure in the lattice version is exact.
 
Following the manipulation, $Z_{DWF}$ is transformed into the following partition function whose action corresponds to the lattice version of $-\mathcal{L}_{N_{b}=2}$ of Eq.~\eqref{eq:dualnf2}:
\begin{equation} 
\begin{split}
Z_{N_{b}=2}[A,B,B';T_{1,2},M] =  & \int Da_{1} Da_{2} Db_{1} Db_{2} Db \\
& Z_{\rm{XY}}[b_{1};T_1]Z_{W}^{*}[a_{1}-b_{1};M,0]Z_{W}[a_{1}-b-B; M, 0] \\
& Z_{\rm{XY}}[b_{2};T_2]Z_{W}[a_{2}-b_{2};M,0]Z_{W}^{*}[a_{2}-b-B'; M, 0] \\
&Z_{BF}[A,B+B',b;M]
\end{split}
\end{equation}
The expected IR behavior for $Z_{N_{b}=2}[X,Y;T,M]$ is relatively straightforward. Based on Sec.~\ref{sec:nuconf}, $Z_{W}^{*}[a_{1}-b_{1};M,0]Z_{W}[a_{1}-b-B; M, 0]$ and $Z_{W}[a_{2}-b_{2};M,0]Z_{W}^{*}[a_{2}-b-B'; M, 0]$ are expected to implement $\delta(b_{1}-b-B)$ and $\delta(b_{2}-b-B')$ under properly tuning $M$. Thus, IR behavior of the second line and the third line of the above equation is simply that of $XY$ models coupled to the gauge field $b+B$ and $b+B'$ respectively. Then, it is natural to expect that $Z_{N_{b}=2}[X,Y;T,M]$ and $-\mathcal{L}_{N_{b}=2}[B,B']$ have the same IR behavior.

The same manipulation transforms $Z_{DF}$ and $\tilde{Z}_{DF}$ into 
\begin{equation}
\begin{split}
Z_{N_{f}=2}[B,B';M'_{1,2},U'_{1,2},M] = & \int Da_{1} Da_{2} Db \, Z_{W}^{*}[a_{1};M'_1,U'_1] Z_{W}[a_{2};M'_2,U'_2] \\
& \hspace{2.7cm} Z_{QF}[A, a_{1},a_{2},B,B';M] \\
\tilde{Z}_{N_{f}=2}[B,B';M'_{1,2},U'_{1,2},M] = & \int D\tilde{a}_{1} D\tilde{a}_{2}Db \, Z_{W}[-\tilde{a}_{1};M'_1,U'_1]Z_{W}^{*}[-\tilde{a}_{2};M'_2,U'_2] \\
& \hspace{2.7cm} \tilde{Z}_{QF}[A,\tilde{a}_{1},\tilde{a}_{2},B,B';M]
\end{split}
\end{equation}
where we have defined
\begin{equation}
\begin{split}
Z_{QF}[A, a_{1},a_{2},B,B';M] =  \int Db \, & Z_{W}[a_{1}-b-B; M, 0] Z_{W}^{*}[a_{2}-b-B'; M, 0] \\
& Z_{BF}[A,B+B',b;M] \\
\tilde{Z}_{QF}[A,\tilde{a}_{1},\tilde{a}_{2},B,B';M]=  \int Db \, & Z_{W}^{*}[-\tilde{a}_{1}+b+B; M, 0] Z_{W}[-\tilde{a}_{2}+b+B'; M, 0] \\
& Z_{BF}[A,B+B',b;M].
\end{split}
\end{equation}
The role of $Z_{QF}[A, a_{1},a_{2},B,B';M]$ and $\tilde{Z}_{QF}[A,a_{1},a_{2},B,B';M]$ is that, upon tuning $M$ properly and interpreting each heavy fermion as implementing level $\pm 1$ Chern-Simons term, we expect
\begin{equation}
\begin{split}
Z_{QF}[A, a_{1},a_{2},B,B';M] \xrightarrow[]{\text{IR limit}} & \int Db \ e^{\frac{i}{2\pi} b d (a_{2}-a_{1}+2B) + \frac{i}{4\pi} (a_{1}-B) d (a_{1}-B) - \frac{i}{4\pi} (a_{2}-B') d (a_{2}-B') }\\ 
& = \delta(a_{2}-a_{1}+2B) \ e^{\frac{i}{4\pi} (a_{1}-B) d (a_{1}-B) -  \frac{i}{4\pi} (a_{2}-B') d (a_{2}-B')} \\
\tilde{Z}_{QF}[A,\tilde{a}_{1},\tilde{a}_{2},B,B';M] \xrightarrow[]{\text{IR limit}} & \int Db \ e^{\frac{i}{2\pi} b d (\tilde{a}_{1}-\tilde{a}_{2}+2B') -\frac{i}{4\pi} (\tilde{a}_{1} -B) d (\tilde{a}_{1} - B) +  \frac{i}{4\pi} (\tilde{a}_{2}-B') d (\tilde{a}_{2}-B') }\\ 
& = \delta(\tilde{a}_{1}-\tilde{a}_{2}+2B') \ e^{-\frac{i}{4\pi} (\tilde{a}_{1} -B) d (\tilde{a}_{1} - B) +  \frac{i}{4\pi} (\tilde{a}_{2}-B') d (\tilde{a}_{2}-B')}
\end{split}
\end{equation}
The key consistency check for the above proposal for the IR limit is whether delta-functions are implemented properly, presumably due to condensing of gauge-invariant composite objects and the subsequent Higgsing of gauge field coupled to the composite objects. This scenario is analogous to the one we studied in Sec.~\ref{sec:nuconf}; we will see that a similar condensation of composite objects occur in the theory $\tilde{Z}_{QF}$ and $Z_{QF}$ occur through the sign-free Monte-Carlo simulation; we present the details in Sec.~\ref{sec:nf2nc}.

\subsection{Emergence of $SU(2)\times SU(2)$ Symmetry}

One remarkable feature of the duality Eq.~\eqref{eq:dualnf2} that lead to its application in deconfined quantum criticality \cite{Wang2017} is the emergence of a global $SU(2)\times SU(2)$ symmetry. Consider the fermion theories $-\mathcal{L}_{N_f=2}$ and $-\tilde{\mathcal{L}}_{N_f=2}$ in Eq.~\eqref{eq:dualnf2} at vanishing background fields $B=B'=0$. Recall that $-\mathcal{L}_{\rm Dirac}^{-, m}[a] = -\mathcal{L}_{\rm Dirac}^{+, m}[a] - \frac{i}{4\pi} ada$. Thus, at $m_2=-m_1$, the theory $-\mathcal{L}_{N_f=2}$ has a manifest $SU(2)_\psi$ symmetry acting on the fermion doublet $(\psi_1, \psi_2)$ of mass $m_2$, and similarly the theory $-\tilde{\mathcal{L}}_{N_f=2}$ has a manifest $SU(2)_\chi$ symmetry acting on $(\chi_1, \chi_2)$ of mass $-m_2$. Recall that in the duality between $-\mathcal{L}_{N_f=2}$ and $-\tilde{\mathcal{L}}_{N_f=2}$, the light fermion in one theory plays the role of a monopole operator (heavy fermion) in the other, so these two $SU(2)$ symmetries are independent. Thus, we have a global $SU(2)_\psi \times SU(2)_\chi$ symmetry in total. This means the boson theory $-\mathcal{L}_{N_b=2}$ also has this symmetry, though not manifest. One may also note that the background fields $B$ and $B'$ can be understood as the $\sigma^3$ gauge fields of the  $SU(2)_\psi$ and $SU(2)_\chi$ symmetry respectively.

We recall that the duality between $-\mathcal{L}_{N_f=2}$ and $-\tilde{\mathcal{L}}_{N_f=2}$ can be regarded as a \emph{self-duality} when $m_1=m_2$. This means at the $m_1=m_2=0$ critical point, the action of the $SU(2)_\psi \times SU(2)_\chi$ symmetry is self-dual. Indeed, as commented above, when $m_2=-m_1$, the two $SU(2)$'s act on fermion doublets of mass $\pm m_2$ respectively, which becomes the same only at zero mass.

In any UV completion, there shall be two UV parameters that play the roles of $m_1$ and $m_2$ in the IR. Then one might hope the UV completion could be such that, if we tune these two UV parameters to satisfy one relation (corresponding to $m_1=m_2$ in the IR) the UV theory would have exact self-duality, and if we tune them to satisfy another relation  (corresponding to $m_1=m_2$ in the IR) the UV theory would have an exact $SU(2)_\psi \times SU(2)_\chi$ symmetry. Unfortunately, this would be hard to simultaneously realize on the lattice, because in the above, we have used the relation $-\mathcal{L}_{\rm Dirac}^{-, m}[a] = -\mathcal{L}_{\rm Dirac}^{+, m}[a] - \frac{i}{4\pi} ada$ in the continuum, which becomes non-exact on the lattice as we have seen in Sec.~\ref{sec:comment_on_BV}. Instead, in our construction in Sec.\ref{sec:nf2}, the \emph{self-dual property is exact at all scales} if we set $T_1=T_2$; on the other hand, \emph{the $SU(2)_\psi \times SU(2)_\chi$ symmetry emerges in the} IR if we make $T_{1,2}$ so that their IR mass $\sim M'_{1,2}-3$ become opposite. The two properties are compatible at the critical point $T_1=T_2=T_c$.

\subsection{Numerical Confirmation of the Implementation of ``Lagrange Multiplier''}
\label{sec:nf2nc}

In this section, we provide the numerical evidence for condensation of the gauge-invariant four-fermion operators in the theory $\tilde{Z}_{QF}[A,a_{1},a_{2},B,B';M]$. This four-fermion operator is charged under $a_{1}-a_{2}+2B'$; condensation of this operator indicates that $\tilde{Z}_{QF}[A,a_{1},a_{2},B,B';M]$ implements $\delta(a_{1}-a_{2}+2B')$, at least for IR modes of the gauge fields. We emphasize that $\tilde{Z}_{QF}$ does not represent $N_{f}=2$ dualities and as in Sec.~\ref{sec:nuconf}, our primary focus in nuemrics is to show emergence of U$(1)$-breaking gapped phase across some parameters rather than investigating nature of a possible critical point.
 

 First, we observe that  the theory $\tilde{Z}_{QF}[A,a_{1},a_{2},B,B';M]$ can be simulated without sign problem when $a_{1}=a_{2}=B=B'=0$, similar to what we did for $Z_{DF}$ investigated in Sec.~\ref{sec:nuconf}, since fermion determinants once again come in complex conjugate pairs. To see this, we observe that $\tilde{Z}_{QF}[A,a_{1},a_{2},B,B';M]$ can be written as:
\begin{equation}
\begin{split}
\tilde{Z}_{QF}[A,a_{1},a_{2},B,B';M] &= Z_{W}^{*}[A+B+B';M,0]Z_{W}[A;M,0] \int Db D \bar{\psi} D \psi D \bar{\chi} D \chi D \bar{\psi}' D \psi' D \bar{\chi}' D \chi'  \\
& e^{\vec{\bar{\psi}}^{T}N(b,-A-B-B',M)\vec{\psi} + \vec{\bar{\chi}}^{T} N^{*}(b,-A,M) \vec{\chi} + \vec{\bar{\psi}'}^{T}N(b,a_{2}-B',M)\vec{\psi'}+ \vec{\bar{\chi}'}^{T} N^{*}(b,a_{1}-B,M)\vec{\chi}}
\end{split}
\end{equation}
Upon excluding $Z_{W}^{*}[A+B+B';M,0]Z_{W}[A;M,0]$ which only contributes to IR physics by providing background topological terms, one can explicitly see that fermion determinants for each configuration of $b$ explicitly comes in complex conjugate pairs when assuming $a_{1}=a_{2}=B=B'=0$.
 
 Now, let us state what observables what we are interested in computing. First, we define the following gauge-invariant two-fermion composite operators:
\begin{equation}
\begin{split}
&\phi_{n,0} = \frac{1}{\sqrt{2}}(\psi_{n,\uparrow}\chi_{n,\uparrow} + \psi_{n,\downarrow}\chi_{n,\downarrow}), \, \bar{\phi}_{n,0} = \frac{1}{\sqrt{2}}(\bar{\psi}_{n,\uparrow}\bar{\chi}_{n,\uparrow} + \bar{\psi}_{n,\downarrow}\bar{\chi}_{n,\downarrow})  \\
&\phi'_{n,0} = \frac{1}{\sqrt{2}}(\psi'_{n,\uparrow}\chi'_{n,\uparrow} + \psi'_{n,\downarrow}\chi'_{n,\downarrow}), \, \bar{\phi}'_{n,0} = \frac{1}{\sqrt{2}}(\bar{\psi}'_{n,\uparrow}\bar{\chi}'_{n,\uparrow} + \bar{\psi}'_{n,\downarrow}\bar{\chi}'_{n,\downarrow}).
\end{split}
\end{equation}
One can see that \textit{the four fermion composite operator} $\phi_{n,0}\phi_{n,0}'$ is coupled to the gauge field $a_{2}-a_{1}+2B'$. While we do not have analytical intuition why this object should condense from expanding Grassmann variables as in Sec.~\ref{sec:nuconf}, we will present numerical results consistent with long-range order associated with $\phi_{n,0}\phi_{n,0}'$, and thus $\tilde{Z}_{QF}$ implements the delta function we desire. Another important consistency check for whether $\tilde{Z}_{QF}$ implements the correct delta functions is that \textit{$\phi_{n,0}$ and $\phi_{n,0}'$ should not condense individually}; if $\phi_{n,0}'$ and $\phi_{n,0}$ condenses as well, $\tilde{Z}_{QF}$ implements two delta functions instead of one and allows one to integrate out one more gauge fields in IR, which is not an expected behavior in our $N_{f}=2$ construction.

The most direct probe of long-range orders associated with $\phi_{n,0}\phi_{n,0}'$ and $\phi_{n,0}$ is the spatial dependence of the following correlators:
\begin{equation}
\begin{split}
C_{4}(n_{1},n_{2}) &= \left\langle \left(\phi_{n_{1},0}\phi'_{n_{1},0}\right) \: \left(\phi_{n_{2},0}\phi'_{n_{2},0} \right) \right\rangle \\
C_{2}(n_{1},n_{2}) &= \left\langle \phi_{n_{1},0} \: \phi_{n_{2},0} \right\rangle
\end{split} 
\end{equation}
Certain observables in this Monte-Carlo simulation converge much slowly compared to the simulation presented in Sec.~\ref{sec:nuconf}, due to the more complicated dynamics. Particularly, we found that it is very difficult to get reasonable estimate for the $C_{4}(n_{1},n_{2})$. However, ``pair susceptibilities" written below can be computed in reasonable amount of time and hence will be used to probe long-range orders:
\begin{equation}
\label{eq:defchi}
\chi_{2F} = \frac{1}{V^2}\sum_{n_{1},n_{2}} C_{2}(n_{1},n_{2}), \qquad \chi_{4F} = \frac{1}{V^2}\sum_{n_{1},n_{2}} C_{4}(n_{1},n_{2})
\end{equation}
Also, we found that adding a Maxwell term for the gauge field $b$ improves the speed of convergence, though its effects on the numerical results are minor; the effect being rather minor is expected, because the Maxwell term is irrelevant. \footnote{For convenience we used a ``compact'' Maxwell, $-S=(1/2g^2) \sum_\Box \cos \left(\vec\triangle\times \vec b\right)_\Box$; see footnote \ref{compact_Maxwell}. We also expect the use of non-compact Maxwell \eqref{eq:Maxwell} would not change this result substantially.}

\begin{figure}
\begin{minipage}{0.5\linewidth}
\includegraphics[width = 1.0\linewidth]{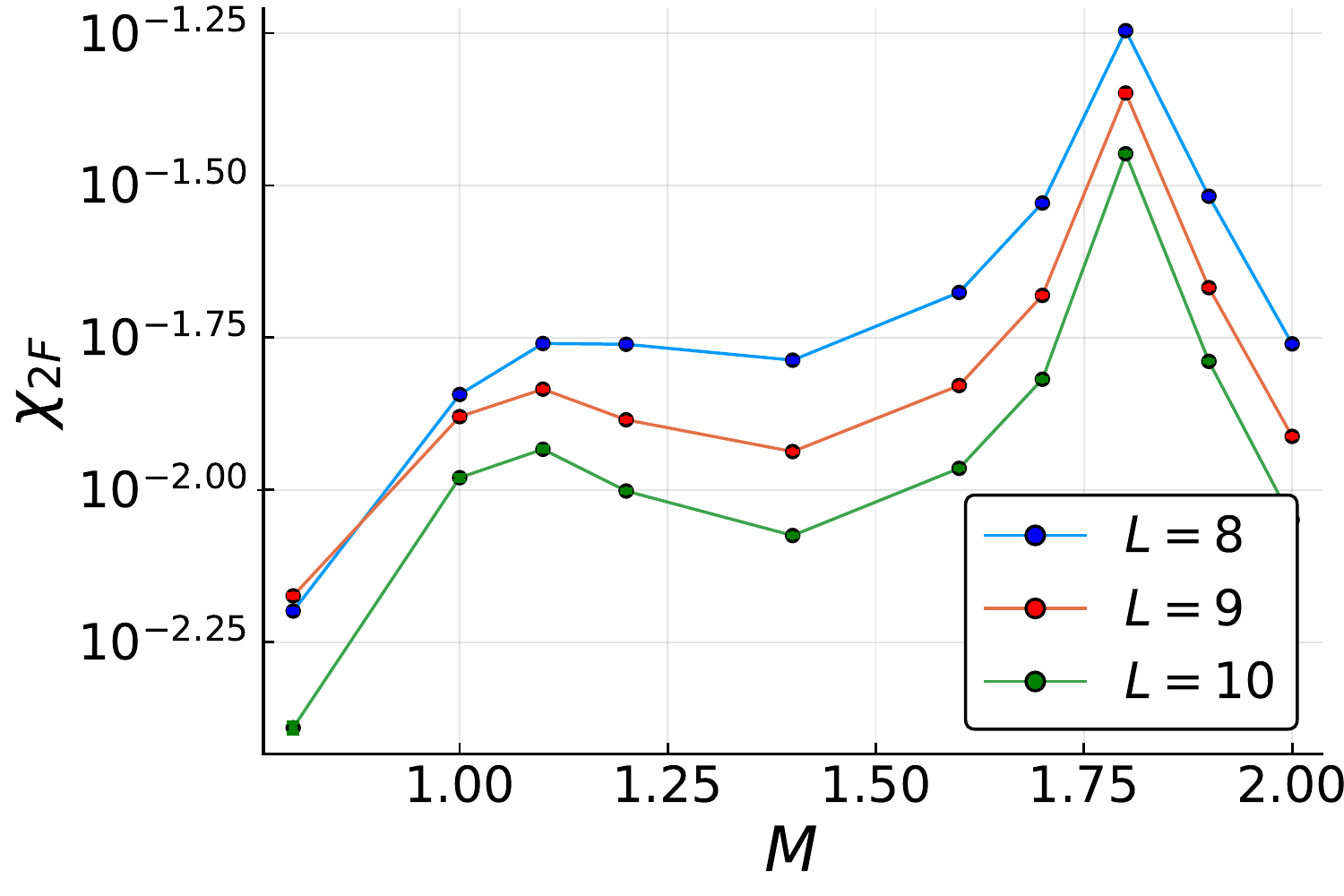}
\end{minipage}
\begin{minipage}{0.5\linewidth}
\includegraphics[width = 1.0\linewidth]{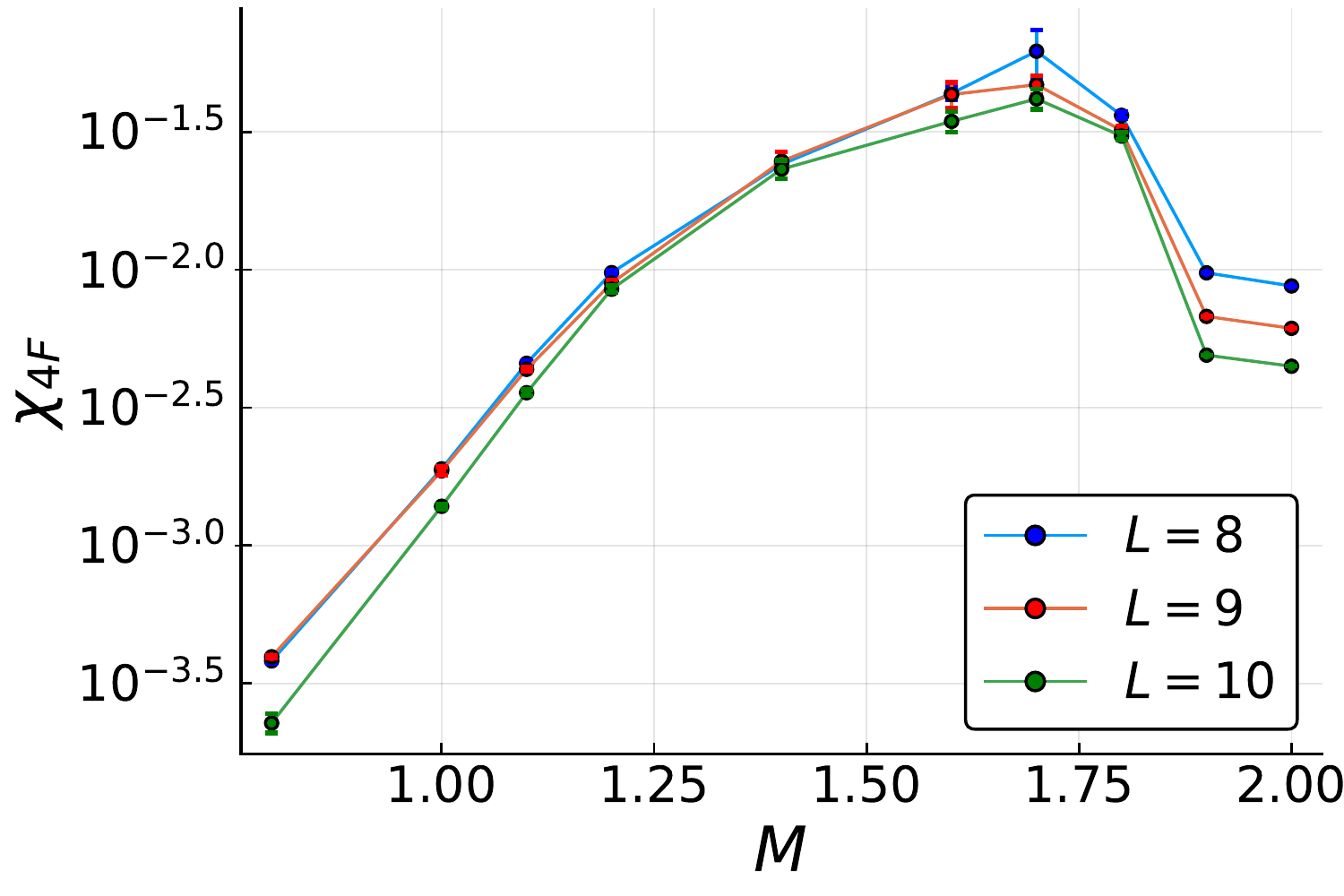}
\end{minipage}
\caption{The plots for $\chi_{2F}$ (left panel) and $\chi_{4F}$ (right panel) versus $M$, different colors corresponding to different system sizes.}
\label{fig:susgraph1}
\end{figure}
 
 We computed $\chi_{2F}$ and $\chi_{4F}$ for $M = 0.8 - 2.0$, setting the inverse Maxwell term strength $g^2 = 100$, for $L \times L \times L$ cubes with $L=8,9,10$;  the computed values are plotted in Fig.~\ref{fig:susgraph1}. On the right panel of Fig.~\ref{fig:susgraph1}, one see a strong peak in $\chi_{2F}$ centered around $M\approx 1.8$; a much weaker peak, more manifest in larger system sizes, can be observed around $M\approx 1.1$. The computed values of $\chi_{2F}$ do not jump otherwise, suggesting that except at those two putative critical points, $\phi_{n,0}$ remains gapped. Meanwhile, as for computed $\chi_{4F}$'s plotted on the right panel, there is a gradual increase of $\chi_{4F}$'s as one tunes from $M=0.8$ to $M=1.4$, followed by a sharp decrease around $M\approx 1.8$. This also suggest the phase transition around $M\approx 1.1$ and $M\approx 1.8$; particularly, the sharp drop at $M\approx 1.8$ in $\chi_{4F}$ suggests that the phase transition is associated with condensation/localization of $\phi_{n,0}\phi_{n,0}'$.
 

The finite-size scaling of $\chi_{2F}$ and $\chi_{4F}$ reveals more information about the nature of the phase. First, let us comment on how $\chi_{4F}$ and $\chi_{2F}$ is expected to scale as one increases the volume. We note that in the thermodynamic limit, one may approximate the sum in Eq.~\eqref{eq:defchi} by integrals. Then, as one approaches thermodynamic limit:
\begin{equation}
\chi_{2F} \sim \frac{1}{V} \int d^{3} r C_{2}(|\vec{r}|), \qquad \chi_{4F} \sim \frac{1}{V} \int d^{3} r C_{4}(|\vec{r}|)
\end{equation}
The above integral allows one to infer correlation function behaviors as the following:
\begin{enumerate}
\item When the correlator decays exponentially or with power law with exponent larger than 3, the integral converges to constant value at thermodynamic limit. Thus, in this case, the corresponding ``susceptibility' should scale like $\frac{1}{V}$. 
\item If the correlator has long-range order, $\chi_{2F}$ and $\chi_{4F}$ should converge to a constant in thermodynamic limit.
\item If the correlator decays with power law with exponent $\alpha < 3$, by similarly using integral one can argue that the corresponding susceptibility should scale like $\frac{1}{V^{\alpha/3}}$. 
\end{enumerate}

As for $\chi_{2F}$, at $M = 1.4-1.7$ and $M =1.9,2.0$, the scenario 1 prevails. Strictly speaking, it is also possible that the correlator decays in power-law with large exponent, yet we conjecture $\phi_{n,0}$ is truly gapped in this regime. Meanwhile, at $M = 0.8$, we cannot observe well-defined scaling behavior of $\chi_{2F}$ - presumably, one needs to study larger systems to extract the correct scaling behavior at this mass value. For $M$ values near $M\approx 1.1$ and $M\approx 1.8$, scenario 2 prevails, suggesting second-order critical points around this Mass values.
 
 As for the behavior of $\chi_{4F}$, at $M = 1.9, 2.0$, scenario 1 prevails once more, indicating $\phi_{n,0}\phi'_{n,0}$ is also gapped in this parameter. Hence, we conjecture that the phase in $M>1.8$ is a trivial gapped phase adiabatically connected to $ M =\infty$ without phase transition. At all other values, scenario 2 or 3 prevails, indicating that $C_{4F}$ is at least power-law decaying in the rest of the $M$- value. Our particular interest is in $M$ value between $1.1$ and $1.8$, the range we conjecture for $\phi_{n,0}\phi'_{n,0}$ condense. However, it is not entirely clear whether scenario 2 or 3 is valid, due to the size of error bar. 
 
\begin{figure}
\begin{minipage}{0.5\linewidth}
\includegraphics[width = 1.0\linewidth]{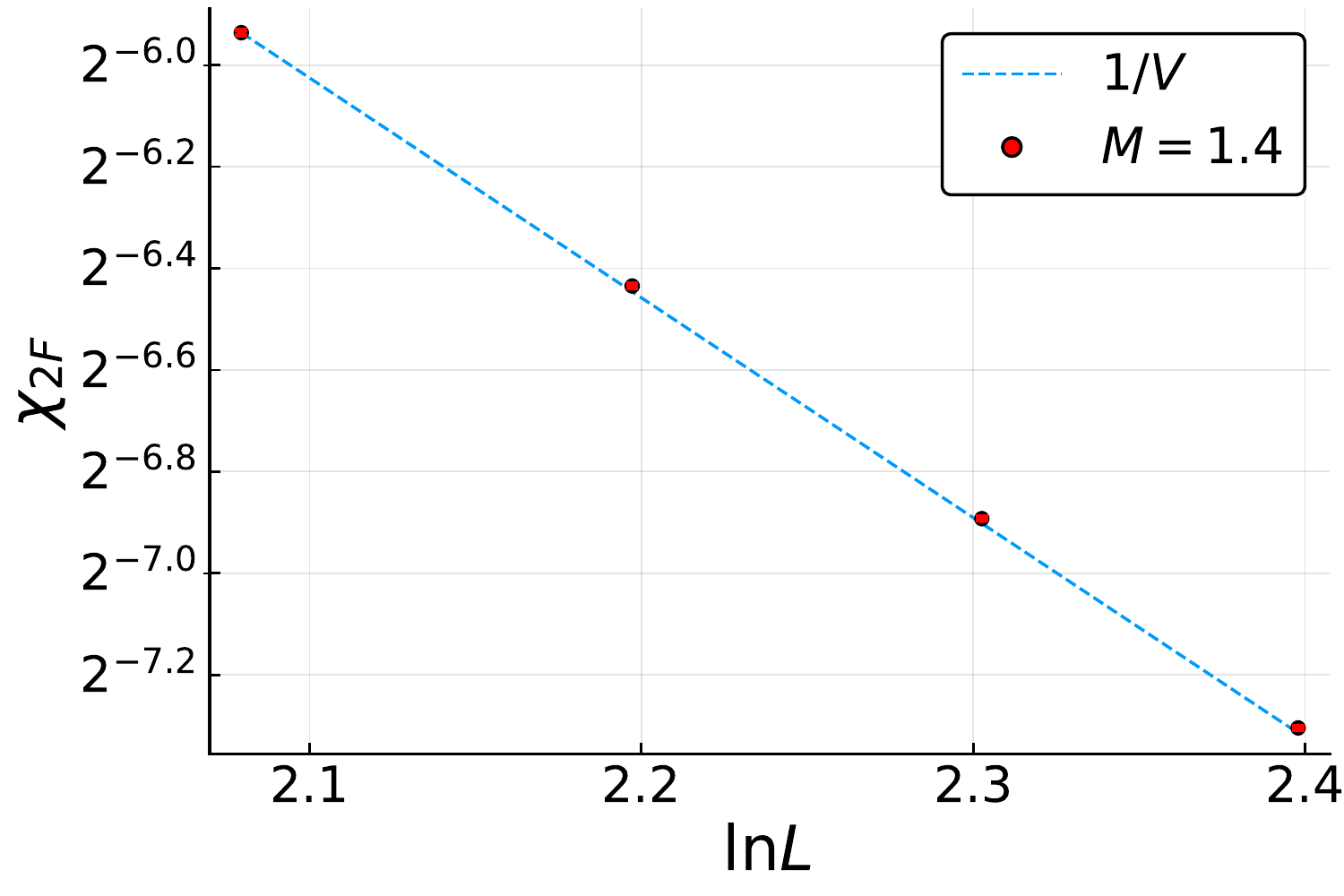}
\end{minipage}
\begin{minipage}{0.5\linewidth}
\includegraphics[width = 1.0\linewidth]{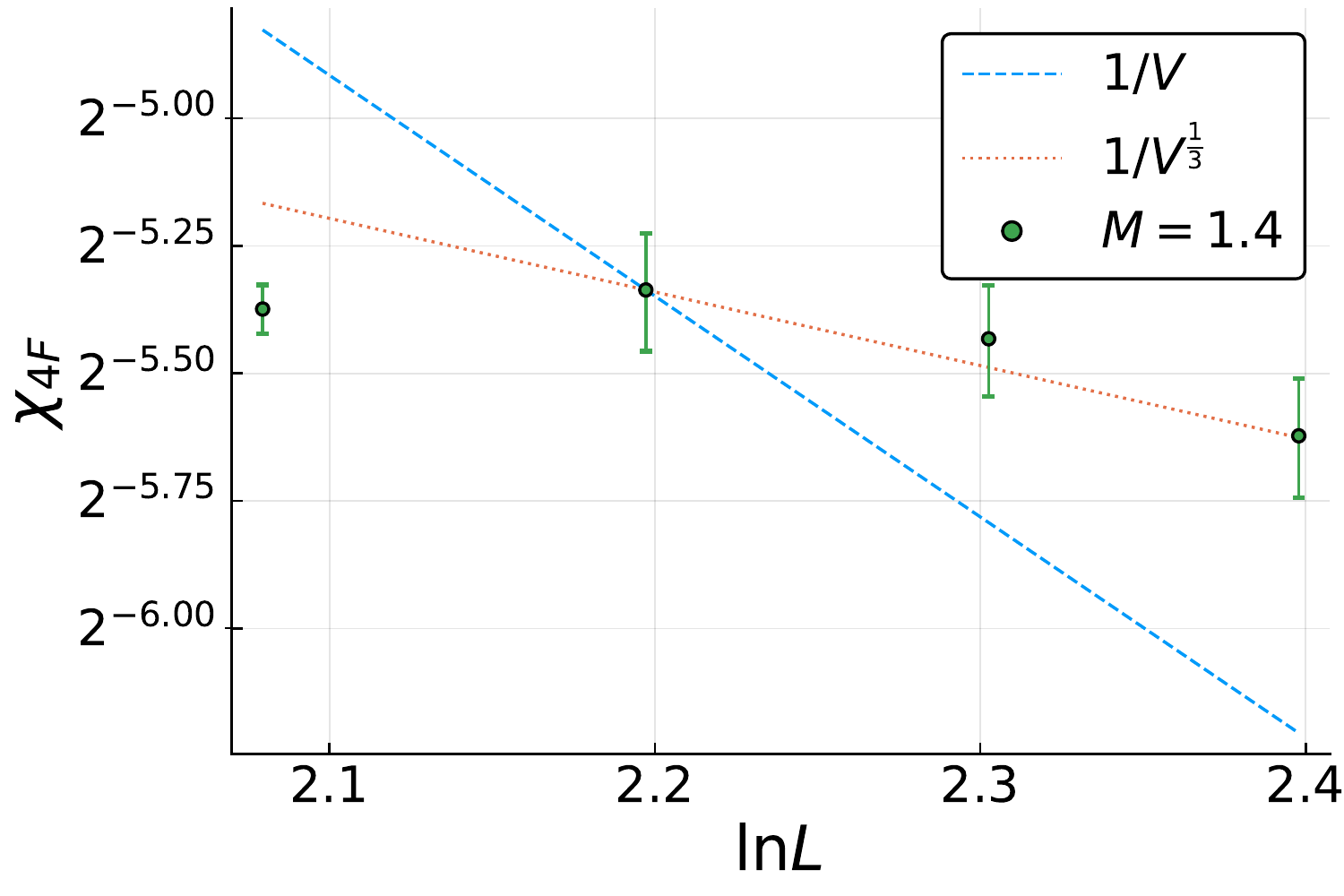}
\end{minipage}
\caption{The plots for $\chi_{2F}$ (left panel) and $\chi_{4F}$ (right panel) for $M=1.4$, $L = 8,9,10,11$. On the left panel, we also plotted blue dashed lines obtained assuming that the computed $\chi_{2F}$ at $L=8$ is exact, and $\chi_{2F}$ exactly scales as $1/V$, for guidance; in a similar spirit, on the right panel we plotted red dashed lines and blue dashed lines which assume that the computed $\chi_{4F}$ at $L=9$ is exact and $\chi_{4F}$ scales as $1/V^{\frac{1}{3}}$ and $1/V$ respectively.}
\label{fig:susgraph2}
\end{figure}
 To study finite-size scaling behavior of $\chi_{2F}$ and $\chi_{4F}$ of the phase sandwiched between two critical points at $M \approx 1.1$ and $M \approx 1.8$ more carefully, we computed $\chi_{2F}$ and $\chi_{4F}$ at $M = 1.4$, $g^2 = 100$ for $L=11$ and plotted them along with the values for $L=8,9,10$ in Fig.~\ref{fig:susgraph2}. One can see clearly on the left panel that $\chi_{2F}$ almost precisely scales like $1/V$, confirming our conjecture that scenario 1 prevails for $\chi_{2F}$ at this $M$ value. The right panel confirms that the computed $\chi_{4F}$ values are fundamentally inconsistent with scenario 1. The red dotted line seems to fit well with data when the value from $L=8$ is excluded, which seems to indicate scenario 3 prevails. However, we note that error bars is are still large here,and while it is clear that $\phi_{n,0}\phi_{n,0}'$ is not gapped, the true nature of scaling behavior remains inconclusive. We believe it is possible that the slow decrease in $\chi_{4F}$ as one increases the system size is non-universal finite-size effect that should go away in the thermodynamic limit, and eventually $\chi_{4F}$ converges to a constant. This conjecture should be verified more carefully in future works.

\section{Non-Relativistic Limit}
\label{sec:nrlimit}
Soon after the original boson-vortex duality was introduced, non-relativistic generalizations of the duality transformation were developed \cite{Dasgupta1981,Fisher1989}. Such generalizations played an important role in the condensed matter realizations of such dualities, enabling, for instance, a more global understanding of the superfluid-Mott insulator quantum phase transitions \cite{Fisher1989a}. On the other hand, the non-relativistic limit of the bosonization and other dualities discussed in this paper can be related to the traditional practice of flux attachment, which is also an important method in condensed matter physics, especially in the quantum Hall effect \cite{zhang1989,lopez1991}.

Let's start with the non-relativistic limit of the free Dirac theory in the continuum, that appeared in Eq.~\eqref{eq:bfdual}. The non-relativistic limit occurs at finite $m>0$, and the fermions are under a chemical potential $\eta$ such that $\eta \gtrsim m$; the range of energy scale of interest is from some $\mu\sim m$ to $\mu+\Delta\mu$, where $\Delta\mu \sim \eta-m \ll m$. The chemical potential makes appearance in the Euclidean signature such that $A=(A_x, A_y, A_z)$ in the first term of $-\mathcal{L}_{\rm Dirac}^{+, m}[A]$ becomes $(A_x, A_y, A_z+i\eta)$; in the Lorentzian signature, $iA_z \rightarrow A_t$, so $\eta$ appears in the Dirac Lagrangian in the form $A_t-\eta$ as expected. We emphasize that this shift of $A_z$ by $i\eta$ is \emph{not} made in the parity anomaly term; this should be taken as part of what we mean by the improperly quantized level-$1/2$ Chern-Simons term.

One may then ask how the chemical potential $\eta$ makes its appearance on the boson side of Eq.~\eqref{eq:bfdual}. There are two ways to interpret $\eta$ on the boson side. To present the simpler interpretation, let's first shift $b$ in Eq.~\eqref{eq:bfdual} by $A$:
\begin{equation}
-\mathcal{L}_{{\rm boson}}[A] =  -|(\partial_\mu-ib'_\mu-iA_\mu) \phi|^2 - r|\phi|^2 - \lambda |\phi|^4 + \frac{i}{4\pi} b' d b'
\end{equation}
and then replace $A_z$ with $A_z + i\eta$ (Euclidean signature) or $A_t$ with $A_t-\eta$ (Lorentzian signature) as in the fermion theory. This means in the bosonic theory, the boson charge density is subjected to a chemical potential $\eta$, as is the fermionic charge density in the fermion theory. Alternatively, we may interpret $\eta$ in the bosonic theory without shifting $b$ to $b'$. We directly shift $A_z$ to $A_z+i\eta$ in the Chern-Simons term in Eq.~\eqref{eq:bfdual}. While the background $dA$ flux can be assumed exact for simplicity, the dynamical $db$ flux might not be, thereby yielding a term $-\eta \int dz \int dx dy (db)_{xy}/(2\pi)$, where $\int dx dy (db)_{xy}/(2\pi)$ is an integer, the total monopole charge over the $x,y$-directions. \footnote{One may wonder why the denominator is $2\pi$ though naively it seems to be $4\pi$. This has to do with the fact that the expression $\frac{i}{4\pi}\int b db$ is incomplete and cannot be taken literally when non-exact monopole charge is present. While we will not introduce the complete expression, one way to convince oneself about the coefficient is that $db/2\pi$ is the current coupled to $A$, so its density component should be $(db)_{xy}/2\pi$.} This term can be understood as a chemical potential for the monopole charge density. The two interpretations of $\eta$ in the bosonic theory are equivalent because the duality states that the boson $\phi$ and the $db$ monopole form a composite object, which is the fermion $\psi$ in the fermionic theory, therefore we can attribute the chemical potential either to the boson or to the monopole.

The two ways of interpreting the chemical potential is obvious in our lattice construction. Let's define $Z_{\rm{XY}}^{\rm{NR}}[b;\eta, T]$ as $Z_{\rm{XY}}[\tilde{b};T]$ where $\tilde{b}$ is to replace $b_{nz}$ on the $nz$ links with $b_{nz} + i\eta$. Likewise, we define $Z_{W}^{\rm{NR}}[A;M,\eta,U]$ as $Z_{W}[\tilde{A};M,U]$, where $\tilde{A}$ is to replace $A_{nz}$ on the $nz$ links with $A_{nz} + i\eta$. Then, our lattice construction for the bosonization duality \cite{Chen2018} may be easily generalized to the mapping between the two partition functions with chemical potential:
\begin{equation}
\begin{split}
&\overline{Z}_{B}^{\rm{NR}}[A;\eta_{1}, \eta_{2}, T,M,U] = \int Db \, Z_{\rm{XY}}^{\rm{NR}}[b;\eta_{1}, T] \: Z_{W}^{\rm{NR}}[A-b;M,\eta_{2},U]\\
&Z_{W}^{\rm{NR}}[A;M',\eta_{1}+\eta_{2},U'].
\end{split}
\end{equation}
Here $\eta_1$ may be viewed as the chemical potential for the boson, $\eta_2$ that for the monopole (heavy Dirac fermion), and in the dual theory they together contribute to the chemical potential for the light Dirac fermion, which is the composite object of the boson and the heavy Dirac fermion.

In what follows, we make another comment about the non-relativistic limit, in connection to the traditional flux attachment. In the traditional non-relativistic flux attachment, a boson may be mapped to any odd number of fluxes attached to a fermion, or any even number of fluxes attached to a boson. On the other hand, in the relativistic duality, say Eq.\eqref{eq:fbf}, a relativistic boson may only be mapped to either $+1$ or $-1$ flux attached to a relativistic fermion, but not the versions with more fluxes. \footnote{If we carry out further $\mathcal{S}, \mathcal{T}$ transformations as introduced in Sec.~\ref{sec:more_S_T}, we will only get relativistic dualities like ``a boson with six fluxes attached is dual to a fermion with five fluxes attached'', unlike in the non-relativistic case where a boson with six fluxes attached is just another boson.} Below, we explain why the relativistic duality is so limited compared to the non-relativistic flux attachement; more exactly, we explain why, as we take the non-relativistic limit of the relativistic theories, extra approximate dualities appear. We note that a similar issue appeared in the context of loop models as well \cite{Fradkin:1996xb,Goldman2018}.

In non-relativistic flux attachment, one may view the attachment of multiple fluxes as the following sequence of operations. A boson can be realized as a fermion with $+1$ flux attached. The fermion, in turn, may be realized as another boson with $+1$ flux attached. This new boson, in turn, may be viewed as another fermion with $+1$ flux attached, and so on. (Similarly if we replace $+1$ flux by $-1$.) Therefore the original boson can be realized as any odd / even number of fluxes attached to a fermion / boson. How is the situation different in the relativistic dualities? The right half of Eq.~\eqref{eq:fbf} states that a boson can indeed be realized as a fermion with $+1$ flux attached. However, it is crucial that this fermion is described by $-\mathcal{L}_{\rm Dirac}^{-, -m}$ which has a $-1/2$ parity anomaly term. The fermion described by $-\mathcal{L}_{\rm Dirac}^{-, -m}$ may \emph{only} be viewed as the original boson with $-1$ flux attached (i.e. the time-reversed version of Eq.~\eqref{eq:bfdual1}), but \emph{cannot} be viewed as another boson with $+1$ flux attached (which would be $-\mathcal{L}_{\rm Dirac}^{+, +m}$ with $+1/2$ parity anomaly term), so the sequence of operations stops there as opposed to the non-relativistic case. In other words, the difference compared to the non-relativistic case lies in that, in relativistic theory, $-\mathcal{L}_{\rm Dirac}^{-, -m}$ and $-\mathcal{L}_{\rm Dirac}^{+, +m}$ are distinct. \footnote{One may wonder what if we trade the $-\mathcal{L}_{\rm Dirac}^{-, -m}$ at hand with $-\mathcal{L}_{\rm Dirac}^{+, -m}$ at the cost of an extra level-$1$ Chern-Simons. We have already discussed this manipulation -- it leads to the boson-vortex duality.}

We expect as we approach the non-relativistic limit in the relativistic theory, the distinction between $-\mathcal{L}_{\rm Dirac}^{-, -m}$ and $-\mathcal{L}_{\rm Dirac}^{+, +m}$ diminishes and then we may, at better and better approximation, carry out the non-relativistic procedure of attaching multiple fluxes. The diminishing of the distinction can be easily understood. Consider $-\mathcal{L}_{\rm Dirac}^{+, +m}$ with finite $m>0$ in the non-relativistic limit. The Dirac sea (lower band), along with the parity anomaly term (which, in the lattice UV completion, arises from other ``doubler'' modes in the Dirac sea), contributes a background $AdA$ Chern-Simons at level $-\sgn(m)/2+1/2=0$, i.e. their net contribution is trivial. The upper band, on the other hand, contains modes with Berry curvature $+\frac{1}{2} |m|/E^3$ in the $(p_x, p_y)$ momentum space, where $E=\sqrt{p_x^2+p_y^2+m^2}$; if the upper band has a Fermi surface at chemical potential $\eta\gtrsim m$, integrating the Berry curvature over the Fermi sea leads to a Hall conductivilty of $(1-|m|/\eta)/2$. The case of  $-\mathcal{L}_{\rm Dirac}^{-, -m}$ is similar: the Dirac sea and the parity anomaly term together makes a trivial contribution, while the upper band has Berry curvature $-\frac{1}{2} |m|/E^3$, and, if at chemical potential $-\eta \lesssim -|m|$, Hall conductivity $-(1-|m|/\eta)/2$. In the non-relativistic limit we are interested in a narrow range of energy scale $\Delta\mu \sim |\eta-m| \ll |m|$ near the bottom of the upper band, around which the difference in the Berry curvature or Hall conductivity between $-\mathcal{L}_{\rm Dirac}^{+, +m}$ and $-\mathcal{L}_{\rm Dirac}^{-, -m}$ are suppressed by powers of $\Delta\mu/|m|$. Therefore, as $\Delta\mu$ becomes small, one may approximately replace $-\mathcal{L}_{\rm Dirac}^{-, -m}$ with $-\mathcal{L}_{\rm Dirac}^{+, +m}$, and thus the attachment of multiple fluxes can be approximately carried out, leading to the sequence of operations in the traditional non-relativistic flux attachment.

\section{Conclusion}
\label{sec:conc}

In this work, we extended the lattice field theory construction of the $U(1)$ bosonization duality to other dualities in the duality web. First, we observed that $\mathcal{S}$ and $\mathcal{T}$ transformations can be implemented by adding heavy Wilson fermions and promoting additionally introduced background gauge fields to the quantum fluctuating ones. This allows one to construct the lattice version of the boson-fermion duality and Son's fermion-fermion duality. 
 
The cost of formulating the aforementioned dualities as exact mappings between the lattice field theories is that integrating out gauge fields cannot be done exactly anymore; so one may doubt whether naively implementing modular transformations from heavy Wilson fermions does really make the lattice field theory flow to the fixed point we imagine. Through numerics, we provided evidence that integrating out gauge fields can be valid operations in the IR, evidenced from condensation of properly charged objects. In addition, we could recover non-trivial time-reversal properties of boson-fermion dualities from the lattice construction as well. We pursued a similar idea in constructing the exact lattice  mapping of $N_{f}=2$ self-dual QED. The numerics shows some evidence that condensation of 4-fermion objects induce the correct delta-function structure needed to implement $N_{f}=2$ self-dual QED as well; however, more extensive numerical simulations are required to fully justify this scenario. 

Caveats regarding the UV regularization of the $\mathcal{S}$ transformation defined in Eq.~\eqref{eq:modtrans} are made manifest through the microscopic construction. We particularly emphasize that the seemly innocent relation, $\mathcal{S} \mathcal{S}^{-1} = \mathbf{1}$, does not automatically hold once a Maxwell UV regularization is imposed. For the relation to work in the context of generating the duality web, one must make the additional assumption (which is made implicitly in the literature) that changing the relative strengths between two gauge couplings involved in $\mathcal{S}$ and $\mathcal{S}^{-1}$ over a finite range does not alter the IR physics.

\section*{Acknowledgement}

We thank Yoni Schattner for advice on numerics and Max Zimet for comments on the manuscript. J. H. S and S.R. are supported by the DOE Office of Basic Energy Sciences, contract DEAC02-76SF00515. J.-Y. C. is supported by the Gordon and Betty Moore Foundation's EPiQS Initiative through Grant GBMF4302.

\begin{appendices}

\section{More on the implementation of the Monte-Carlo algorithm}

Here we outline in more details how to implement Monte-carlo simulations utilized in the analysis of Sec.~\ref{sec:nuconf} and Sec.~\ref{sec:nf2nc}. We focus on the theory $Z_{DW}$ in Sec.~\ref{sec:nuconf} -- simulating $\tilde{Z}_{QF}$ can be done with minor modifications which we will specify later.

 First, we note that one can write down the partition function $Z_{DW}[B,b;M_{1},M_{2}]$ in the following form:
\begin{equation}
\begin{split}
Z_{DW}[B,b;M_{1},M_{2}] &= \int Da D\bar{\psi} D\psi D\bar{\chi} D\chi e^{ \vec{\bar{\psi}}^{T}N(a,B,M_{1})\vec{\psi}+ \vec{\bar{\chi}}^{T}N^{*}(a,b,M_{2})\vec{\chi}} \\
&= \int Da \det{N(a,B,M_{1})}\det{N^{*}(a,b,M_{2})}
\end{split}
\end{equation}

Here, $\vec{\psi}$, $\vec{\chi}$, $\vec{\bar{\psi}}^{T}$ and $\vec{\bar{\chi}}^{T}$ are vectors of Grassmann variables; $\vec{\psi}$ are defined by
\begin{equation}
\label{eq:defgrasvec}
\vec{\psi}^{T} = \begin{pmatrix}
\psi_{1,\uparrow}& \psi_{1,\downarrow}& \psi_{2,\uparrow}& \psi_{2,\downarrow} \cdots
\end{pmatrix}
\end{equation}
$\vec{\chi}$, $\vec{\bar{\psi}}^{T}$ and $\vec{\bar{\chi}}^{T}$ are defined likewise. $N(a,B,M_{1})$ is a $2n \times 2n$ matrix that characterize the lattice action. Note that at $B=b=0$ and $M_{1}=M_{2}$, the matrix associated with the part of the action with $\chi$ fermions and the one associated with $\psi$ fermions are \textit{complex conjugate} of each other. Thus, each gauge field configuration effectively has a positive weight $|\det{N(a,0,M)}|^2$, and sum of all these weights define the partition function.  Also, at $B=b=0$ and $M_{1}=M_{2}$, four-point correlation functions as the following:
\begin{equation}
\big\langle  \bar{\psi}_{\alpha_{1}} \psi_{\alpha_{2}} \bar{\chi}_{\beta_{1}} \chi_{\beta_{2}} \big\rangle = \frac{\int Da \quad N_{\alpha_{1}\alpha_{2} }^{-1}(a,0,M)  N_{\beta_{1}\beta_{2}}^{*-1}(a,0,M)  |\det{N(a,0,M)}|^{2}}{\int Da \quad |\det{N(a,0,M)}|^{2}} 
\end{equation}
 In the above equation, $\alpha_{1}$, $\alpha_{2}$, $\beta_{1}$, $\beta_{2}$ denote indices for Grassman variables.
The message we want to deliver from the above equation is that correlation functions can be computed via Monte-Carlo simulation by sampling gauge field configurations so that each gauge field configuration $a$ has a relative probability distribution $|\det{N(a,0,M)}|^{2}$ and taking average of proper matrix elements, in this case $N_{\alpha_{1}\alpha_{2} }^{-1}(a,0,M) N_{\beta_{1}\beta_{2}}^{*-1}(a,0,M)$. One can easily generalize this to higher-point correlation functions -- the matrix elements to be taken average can be readily determined by Wick's theorem. One can similarly proceed for $\tilde{Z}_{QF}$ to see that one can simulate $\tilde{Z}_{QF}$ simply by modifying the weight for each gauge field configuration $a$ to be $|\det{N(a,0,M)}|^{4}$ instead of $|\det{N(a,0,M)}|^{2}$.



 As for pratical implementations, one can employ standard Metropolis algorithm to sample gauge field configurations. For each step of the algorithm, the program proposes a gauge field configuration to be updated from $\{ a \}$ to $\{ a' \}$ (For the purposes in this paper, one can safely restrict the update to be local; that is, the gauge field configuration $\{ a \}$ and $\{ a' \}$ only differs locally). The program accepts the new proposed configuration with the probability $P(\{ a \} \rightarrow \{ a' \})$, defined as:
\begin{equation}
P(\{ a \} \rightarrow \{ a' \}) = \min \left( 1, \frac{|\det{N(a',0,M)}|^2}{|\det{N(a,0,M)}|^2} \right).
\end{equation}
This procedure allows one to sample the gauge field configurations whose probability distribution is proportional to $|\det{N(a,0,M)}|^2$. As for simulating $\tilde{Z}_{QF}$, one can simply change the Metropolis weight $\frac{|\det{N(a',0,M)}|^2}{|\det{N(a,0,M)}|^2}$ to $\frac{|\det{N(a',0,M)}|^4}{|\det{N(a,0,M)}|^4}$.

 We briefly comment on some practical issues on the Monte-Carlo simulation. At each step of the simulation, one may store gauge field configuration $\{ a \}$ and inverse matrix  $N^{-1}(a,0,M)$. This allows one to readily extract matrix elements needed for computing correlation functions from the inverse matrix $N^{-1}(a,0,M)$. Also, storing the inverse matrix in addition to gauge field configuration allows one to the Metropolis probability $\frac{|\det{N(a',0,M)}|^2}{|\det{N(a,0,M)}|^2}$ in $O(1)$ time. To see this, for simplicity, let us assume that we consider an update from ${a}$ to ${a'}$, and ${a}$ and ${a'}$ have a different gauge field value only at one link.\footnote{Generically, for an update scheme that update $n$ gauge fields at once, the Metropolis probability can be computed in $O(n)$ time -- the complexity is completely independent of the size of the system.} Then, $\Delta(a',a,M) = N(a',0,M) - N(a,0,M)$ is a matrix whose elements are zero everywhere except for two $2 \times 2$ blocks. Thus, 
\begin{equation}
\begin{split}
\frac{|\det{N(a',0,M)}|^2}{|\det{N(a,0,M)}|^2} &= |\det{N(a',0,M)N^{-1}(a,0,M)}|^2 \\
&= | \det{(I + \Delta(a',a,M) N^{-1}(a,0,M))}|^2
\end{split}
\end{equation}
 One can show that thanks to sparsity of the matrix $\Delta(a',a,M)$, $\Delta(a',a,M) N^{-1}(a,0,M))$ is a matrix whose non-zero elements lie on only four columns. Also, by performing column operations, one can show that sixteen non-zero elements of $\Delta(a'-a,M) N^{-1}(a,0,M))$ that form $4 \times 4$ block diagonals are the only ones that contribute to non-trivial Metropolis probability; all other non-zero elements can be ignored. Thus, while the Metropolis algorithm involves ratio of determinants of large matrices which are numerically costly to compute, storing the inverse matrix and restricting to local updates allow transition probability to be computed in $O(1)$ time. 
 
 While computing the Metropolis probability can be implemented in a numerically inexpensive way, updating the matrix inverse $N^{-1}(a,0,M)$ to $N^{-1}(a',0,M)$  when the proposed change is accepted is difficult to be done efficiently and hence act as the bottleneck of this Monte-Carlo simulation. When restricted to local updates, $N^{-1}(a',0,M)$ can be computed from $N^{-1}(a,0,M)$ using Shermann-Morrison formula in $O(V^2)$ complexity, $V$ denoting volume of the system we are simulating; this is the most efficient way to update the matrix inverse from our knowledge. While this is numerically costly, this is much faster than computing $N^{-1}(a',0,M)$ from scratch with complexity $O(V^{3})$.
 
 Finally, recall that when investigating spatial dependence of correlation functions, we made one direction(say $z$ direction) longer than other, forming a ``prism" geometry, and investigated how correlators behave on $z$ axis. We note that for this purposes, it is more efficient to compute the following quantities: 
\begin{equation}
C(n_{z}) = \frac{1}{L_{x}L_{y}}\sum_{n_{x},n_{y}} \left\langle \bar{\phi}_{(n_{x},n_{y},0),0} \phi_{(n_{x},n_{y},n_{z}-),0} \right\rangle
\end{equation} 
 $C(n_{z})$ is the ``spatially-averaged version" of the correlation function we intend to study. Our action is translationally invariant, so if one can average over \textit{all} gauge configurations, $C(n_{z}) = \left\langle \bar{\phi}_{(n_{x},n_{y},0),0} \phi_{(n_{x},n_{y},n_{z}-),0} \right\rangle$ for any $n_{x}$ and $n_{y}$. However, in the Monte-Carlo simulation, we do random sampling of gauge fields, and choosing specific gauge field configurations break translational invariance. The above quantity takes average spatially within the same configuration and the system as well, allowing one to obtain more fastly convergent results.
 

\end{appendices}

\bibliography{duality_new}{}
\bibliographystyle{utphys}
\end{document}